\documentclass[aip, rsi, amsmath, amssymb, reprint]{revtex4-1}

\usepackage{graphicx}
\usepackage{dcolumn}
\usepackage{bm}
\usepackage[utf8]{inputenc}	
\usepackage[T1]{fontenc} 
\usepackage{mathptmx}
\usepackage{xcolor} 
\usepackage{hyperref} 
\graphicspath{ {./figures/} }
\begin{document}
	
	\preprint{AIP/123-QED}
	
	\title[Uncovering temporal regularity]{Uncovering temporal regularity in atmospheric dynamics through Hilbert phase analysis}
	
	\author{Dario A. Zappal\`a}
	\affiliation{%
		Universitat Polit\`ecnica de Catalunya, Departament de F\'isica, Rambla St. Nebridi 22, 08222 Terrassa, Barcelona, Spain
	}%
	\author{Marcelo Barreiro}%
	\affiliation{ Instituto de F\'isica, Facultad de Ciencias, Universidad de la Rep\'ublica, Igu\'a 4225, Montevideo, 11400 Uruguay
	}%
	
	\author{Cristina Masoller}
	\email{cristina.masoller@upc.edu.}
	\homepage{http://www.fisica.edu.uy/~cris/.}
	\affiliation{%
		Universitat Polit\`ecnica de Catalunya, Departament de F\'isica, Rambla St. Nebridi 22, 08222 Terrassa, Barcelona, Spain
	}%
	
	\date{\today}
	
	\begin{abstract}
		Uncovering meaningful regularities in complex oscillatory signals is a challenging problem with applications across a wide range of disciplines.
		Here we present a novel approach, based on the Hilbert transform (HT). 
		We show that temporal periodicity can be uncovered by averaging the signal in a moving window of appropriated length, $\tau$, before applying the HT.
		As a case study we investigate global gridded surface air temperature (SAT) datasets.
		By analysing the variation of the mean rotation period, $\overline{T}$, of the Hilbert phase as a function of $\tau$, we discover well-defined plateaus.
		In many geographical regions the plateau corresponds to the expected one-year solar cycle; however, in regions where SAT dynamics is highly irregular, the plateaus reveal non-trivial periodicities, which can be interpreted in terms of climatic phenomena such as El Niño.
		In these regions, we also find that Fourier analysis is unable to detect the periodicity that emerges when $\tau$ increases and gradually washes out SAT variability. 
		The values of $\overline{T}$ obtained for different $\tau$s are then given to a standard machine learning algorithm.
		The results demonstrate that these features are informative and constitute a new approach for SAT time series classification.
		To support these results, we analyse synthetic time series generated with a simple model and confirm that our method extracts information that is fully consistent with our knowledge of the model that generates the data.
		Remarkably, the variation of $\overline{T}$ with $\tau$ in the synthetic data is similar to that observed in real SAT data.
		This suggests that our model contains the basic mechanisms underlying the unveiled periodicities.
		Our results demonstrate that Hilbert analysis combined with temporal averaging is a powerful new tool for discovering hidden temporal regularity in complex oscillatory signals.
	\end{abstract}
	
	\maketitle
	
	\begin{quotation}
		Extracting meaningful information from complex oscillatory signals is a challenging task with applications across disciplines. Here we propose a novel technique, based on Hilbert analysis, and apply it to global observational surface air temperature (SAT) datasets. 
		We show that, by combining moving temporal average with Hilbert analysis, we can uncover underlying regularities in SAT dynamics, which are not always detected by Fourier spectral analysis. Specifically, by changing the length, $\tau$, of the moving temporal average and analysing how the mean rotation period of the Hilbert phase, $\overline{T}$, depends on $\tau$, we discover, in specific geographical regions, well-defined plateaus that reveal hidden periodicity in SAT dynamics. A main advantage of the technique proposed here is that it allows continuous tuning of the time scale in which SAT variability is washed out. Moreover, using a machine learning algorithm, we show that the variation of $\overline{T}$ with $\tau$ can be used for classification of SAT dynamics in different regions. 
	\end{quotation}

	\section{\label{sec:intro}Introduction}
	
	Our climate is an extremely complex, nonlinear system with interacting sub-systems and feedback loops that act at various spatial and temporal scales \cite{dijbook}.
	To advance in the understanding of the climate system, models with different levels of complexity can be used \cite{Trenberth1993,models}.
	While simple models only provide a good understanding of basic phenomena,  state-of-the-art models allow for unprecedented predictability \cite{weather_prediction}; however, the complexity of the models can obscure the interpretation of their predictions.
	
	In the last two decades, the availability of satellite data and advances in data mining \cite{data_availability2} have lead to the development of many data-driven approaches to understand, characterise and predict our climate, directly from the observed data. Standard tools in climate data analysis include Empirical Orthogonal Functions, Fourier and Wavelet methods \cite{eofs,manfred,wavelet}. In recent years alternative approaches, well-known in other disciplines, are being increasingly used because they have been demonstrated to be able to extract meaningful information for climate predictability~\cite{havlin_prediction_ninio, boers_nat_com, veronika_2016, emilio_2016, ncomm_2017, ott_2018, boers_2019}.
	
	The Hilbert transform (HT), which has been used to investigate a wide range of oscillatory signals (physiological, geophysical, neurological, etc.~\cite{huang_1998,palus,bib:Huang2008,neuro,neuro2,mario,bib:ECG}), is a useful tool for climate data analysis~\cite{bib:our-HilbAnalysis,bib:our-ESD} because climatological variables typically have a degree of seasonality due to solar forcing.
	The HT provides, for an oscillatory time series, an analytic signal from which instantaneous amplitude, phase, and frequency can be derived (see Section~\ref{hilb:an}).
	While the Hilbert transform can be applied to any signal, the instantaneous amplitude coincides with the envelope of the signal, and the instantaneous frequency corresponds to the frequency of the maximum of the power spectrum computed in a running window, only if the signal is ``narrow band'' (see Sec. A2.1 in \cite{bib:Pikovsky} and references therein).
	Signals that do not fulfil the ``narrow band'' criterion are often pre-filtered in a narrow frequency band.
	However, in the specific case of surface air temperature (SAT) with daily resolution, we have shown that the Hilbert transform applied to raw, unfiltered SAT time series allows to uncover meaningful information~\cite{videos}, in spite of the fact that SAT time series are not narrow band.
	We unveiled spatially coherent global patterns of high frequency dynamics~\cite{bib:our-HilbAnalysis} and we uncovered the geographical regions that have experienced important changes in SAT dynamics over the last 30 years~\cite{bib:our-ESD}.
	This success of the Hilbert method for extracting relevant information is due to the fact that, in many geographical regions, SAT time series have well-defined periodicity imposed by the annual cycle of solar forcing. 
	
	Using the Hilbert transform, here we propose a novel framework for detecting underlying temporal regularities in complex oscillatory signals. As a case study we consider SAT time series in a grid over the Earth's surface, covering the last 39 years with daily resolution. We show that underlying regularities can be extracted from the Hilbert phase, if SAT is averaged over a moving window of appropriate length, $\tau$, before applying the HT.
	Our approach is based on the analysis of how the mean rotation period, $\overline{T}$, of the Hilbert phase (see \emph{Methods}) depends on $\tau$. We discover that it has a non-trivial dependence, with one or more plateaus that reveal the presence of stochastic periodicity in SAT dynamics, which is uncovered when smoothing SAT time series in a window of $\tau$ days, with $\tau$ in a plateau.
	
	To gain further insight, we analyse how the instantaneous Hilbert phase $\varphi(t)$ depends on the date of the year, for various values of $\tau$. For $\tau$ within a plateau, the plot of $\varphi(t)$ vs.\ date unveils temporal structures, which provide a qualitative way to evaluate differences in SAT dynamics in different regions. We discuss, as particular examples, the phase and SAT dynamics in six geographical regions, which we refer to as regular ($\overline{T}\sim 1$ year, regardless of $\tau$), quasi-regular ($\overline{T}$ reaches a one year plateau, for $\tau$ large enough), double period ($\overline{T}\sim$ half year in a range of values of $\tau$), irregular (in the plot of $\overline{T}$ vs.\ $\tau$ no plateau is found), El Niño ($\overline{T}$ vs.\ $\tau$ displays a plateau at $\overline{T}\sim 4$ years, which is consistent with the El Niño phenomenon) and QBO (because in this site a plateau is found at $\overline{T}\sim 2.5$ years, which is consistent with the Quasi-Biennial Oscillation). 
	
	To further demonstrate that the analysis of the variation of the mean rotation period of the Hilbert phase, $\overline{T}$, with the length of the average window, $\tau$, indeed extracts meaningful information from SAT time series, we use the k-means clustering algorithm to classify the gridded sites of the dataset into distinct geographical regions, based on the variation of $\overline{T}$ with $\tau$. We also compare Hilbert and Fourier analysis and show that the extracted oscillatory component with periodicity $\overline{T}$ is not always detected by Fourier analysis. These differences can be expected as only for narrow band signals the instantaneous frequency corresponds to the frequency of the maximum of the power spectrum computed in a running window \cite{bib:Pikovsky}.

	\section{\label{sec:methods}Methodology}

	\subsection{Datasets}
	ERA-Interim reanalysis~\cite{bib:ERA-Interim} covers the period from 1979 to 2017 with daily time resolution, and spatial resolution of $73 \times 144$ (2.5$^\circ$ in latitude and in longitude).
	Therefore, we have $N = 73 \times 144 = 10512$ SAT time series, each with $L = 14245$ data points (i.e., days).
	NCEP Reanalysis 2~\cite{bib:NCEP-Rean2}, which is used in Appendix~\ref{app:compar} for the purpose of comparison, has daily resolution and a spatial resolution of $192 \times 94$ ($1.875^\circ$ in longitude and approximately $1.9^\circ$ in latitude).
	To indicate the raw SAT time series we use the notation $r_j(t)$, where $j \in [1, N]$ represents the geographical site and $t \in [1, L]$ represents the day.
	
	\subsection{Hilbert analysis}
	\label{hilb:an}
	First, we smooth the time series $r_j(t)$ by taking a temporal average over a moving window of length $\tau$.
	Then, we remove the linear trend and normalise the time series to zero mean and unit variance.
	The smoothed, detrended and normalised time series is referred to as $x_j(t)$.
	The Hilbert transform of $x_j(t)$, $H[x_j](t)=y_j(t)$, allows to define the analytic signal, $h_j(t) = x_j(t) + i y_j(t)$, from which we calculate the \emph{instantaneous phase} time series $\varphi_j(t) = \arctan ({y_j(t)}/{x_j(t)})$.
	Taking into account the individual signs of $x_j(t)$ and $y_j(t)$, we can obtain $\varphi_j(t)$ in the interval $[-\pi, \pi]$.
	The next step is to unwrap the phase series, adding multiples of $2\pi$ at each time step to avoid the sudden jumps between $\pi$ and $-\pi$.
	Because the HT, calculated over a finite time, gives error near the extremes of the series, after the calculations we disregard the initial and final 2 years.
	Thus, we analyse phase time series of 35 years (from 1981 to 2015) with $L = 12783$ data points.
	
	Finally, we calculate the mean rotation period of the Hilbert phase as 
	\begin{equation}
	\overline{T}_j = 2\pi \frac{\Delta t}{\Delta \varphi_j}
	\end{equation}
	where $\Delta t$ is the time interval between the first and the last day of the time series (in units of years) and $\Delta \varphi_j$ (in radians) is the variation of the unwrapped phase during this time interval.
	
	We repeat these calculations for all the odd values of the smoothing length $\tau$ between 1 day (no smoothing) and 149 days.
	We limit the length to this range in order to avoid filtering out the seasonal cycle.
	At the end, we obtain the mean period $\overline{T}$ as a function of the smoothing length $\tau$.
	We will see that analysing this dependence we can unveil temporal regularities of the original SAT series.
	
	\subsection{Analysis of synthetic series}
	\label{analysis:synth}
	To give a first proof of our analysis method, we begin by applying it to synthetic time series generated with a simple model.
	We will show that our analysis returns information which is fully consistent with our knowledge of the equation that generates the series.
	In other words, the variation of $\overline{T}$ with $\tau$ is as one would expect, considering the parameters used to generate the synthetic time series.
	
	The model that we consider is the sum of two sinusoidal oscillations (the annual cycle and a slower oscillation) and autoregressive noise:
	\begin{equation}
	S(t) = \frac{a\sqrt{2}\cos(\omega_\textrm{sl} t) + b\sqrt{2}\cos(\omega_\textrm{y} t) + c \varepsilon(t)}{\sqrt{a^2+b^2+c^2}},
	\label{eq:syntetic}
	\end{equation}
	where $\omega_\textrm{sl} = \frac{1}{4} 2\pi$ rad/year represents a slow 4-year cycle, $\omega_\textrm{y} = 2\pi$ rad/year represents a one-year cycle, and $\varepsilon(t)$ represents AR(1) noise with zero mean and unit variance and with persistence $\gamma \in [0, 1]$.
	We use AR(1) because it is the usual null hypothesis to model climate data \cite{bib:Hasselmann}.
	With this model, using a sampling time $\Delta t =1$ day, we generate time series of the same length as ERA daily reanalysis (14245 days).
	The parameters $a,b,c$ allow us to vary the amplitude of the three components, while the normalization factors ($\sqrt{2}$ and $\sqrt{a^2+b^2+c^2}$) keep constant the first and second moment of the distribution of $S(t)$ values (zero mean and unit variance).
	
	\begin{figure}[tbp] 
		\centering
		\begin{tabular}[t]{rrr}
			\includegraphics[height=0.3\columnwidth]{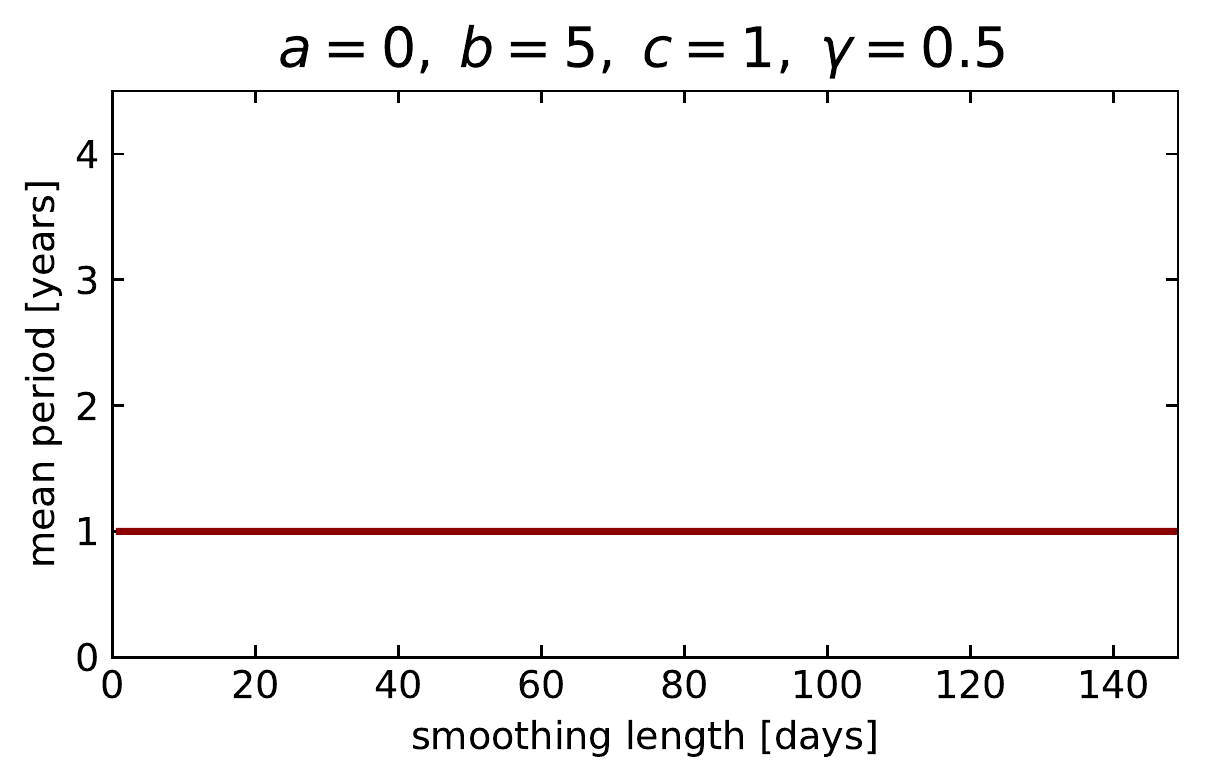}
			&
			\includegraphics[height=0.3\columnwidth]{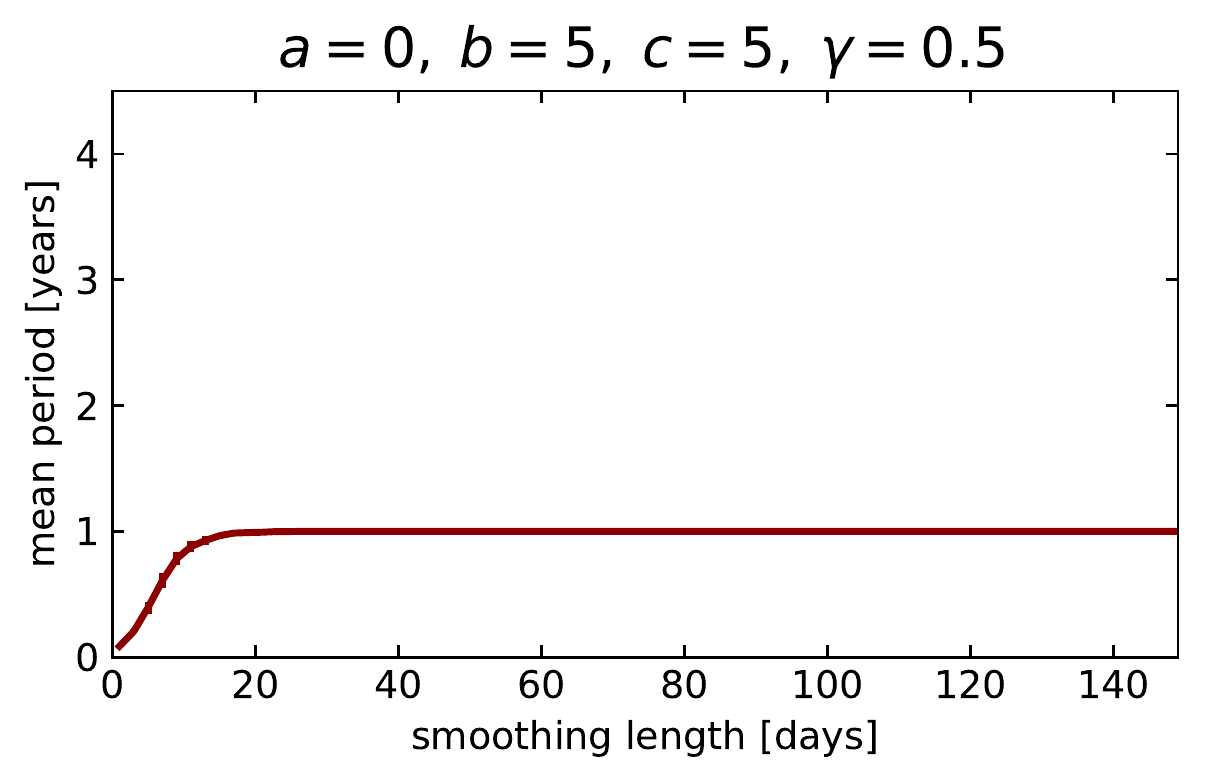}
			\\
			\includegraphics[height=0.3\columnwidth]{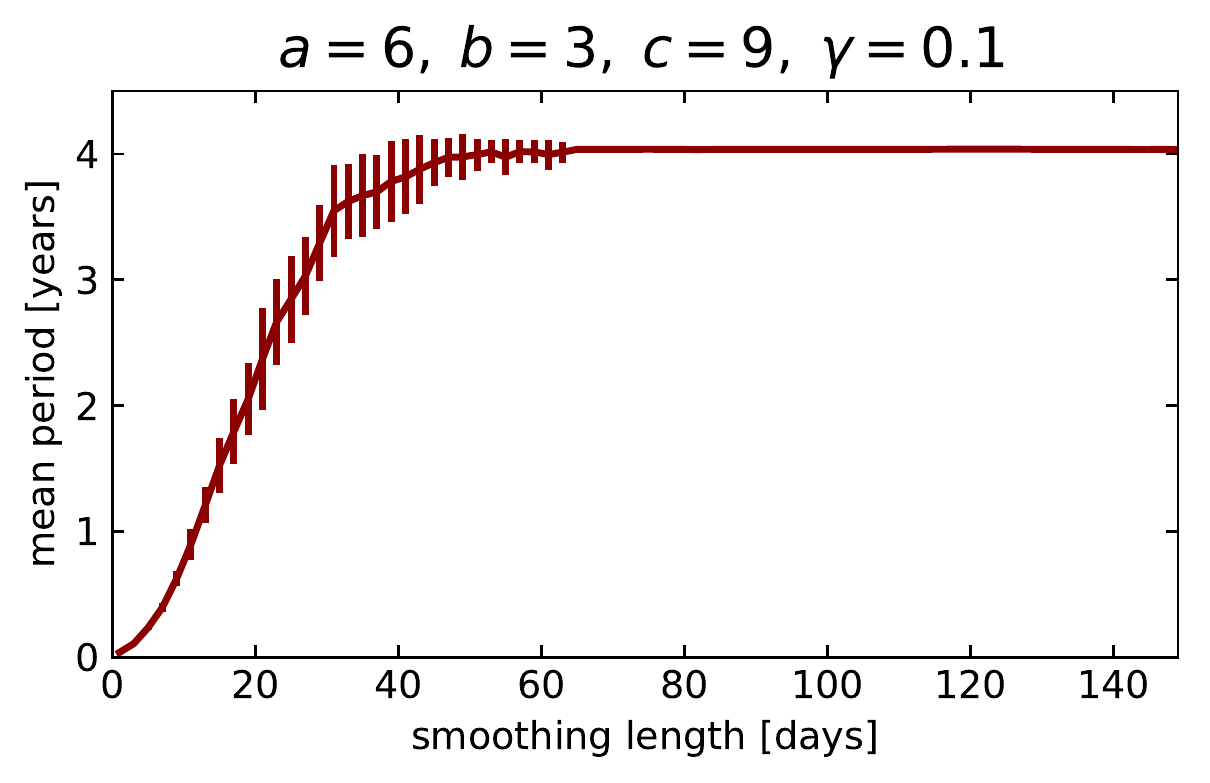}
			&
			\includegraphics[height=0.3\columnwidth]{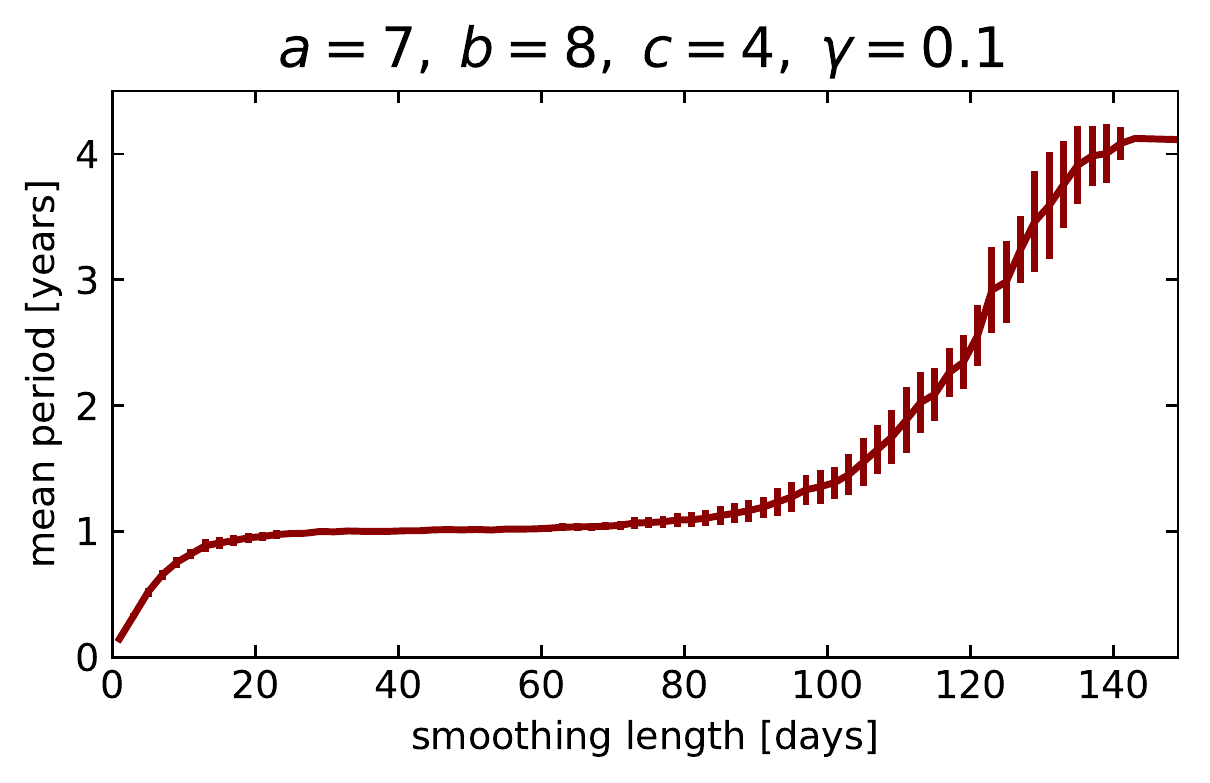}
		\end{tabular}
		\caption{Mean period $\overline{T}$ as a function of the smoothing length $\tau$.
			We show the results from 20 realisations of the synthetic series (the error bars represent the standard deviations).}
		\label{fig:results-synth}
	\end{figure}
	From the synthetic time series we calculate the variation of $\overline{T}$ as a function of $\tau$, following the method that we described in Section~\ref{hilb:an}.
	Figure~\ref{fig:results-synth} displays the results obtained from four different choices of the parameter set.
	For ($a=0, b=5, c=1, \gamma=0.5$), the mean period is always equal to 1 year, regardless of the smoothing length.
	This is an expected result, since these parameters produce a time series dominated by the 1-year oscillation, a circumstance that is not changed by smoothing.
	For ($a=0, b=5, c=5, \gamma=0.5$), the mean period is initially near to 0, but it grows with smoothing and reaches a stable plateau at $\overline{T} = 1$ year.
	This is compatible with the fact that noise has the same weight as the 1-year oscillation, so the fast oscillations produced by noise contribute to the evolution of Hilbert phase and give a fast dynamics.
	When we smooth sufficiently ($\tau \approx 20$ days), we reduce the fast noisy oscillations and leave the 1-year oscillation as the dominating component.
	For ($a=6, b=3, c=9, \gamma=0.1$), again the mean period is initially near to 0, but with smoothing it grows until it reaches a stable plateau at $\overline{T} = 4$ years.
	This is easily explained by the fact that noise has more weight than the regular oscillations, while the 4-year oscillation has more weight than the 1-year oscillation.
	By smoothing, we gradually wash out the fast noisy oscillation modes and we are left with a time series in which the 4-year oscillation dominates on the 1-year oscillation and on the remaining noisy oscillations.
	For ($a=7, b=8, c=4, \gamma=0.1$), the mean period starts with a low value (lower than 1 year), then it increases and reaches a plateau of $\overline{T} = 1$ year.
	Then, it increases again until it finally reaches a plateau at $\overline{T} \approx 4$ years.
	This behaviour is explained by the fact that smoothing initially eliminates the effects of noise on the mean period.
	Since the 1-year cycle has a slightly bigger weight than the 4-year cycle, the former cycle initially prevails, giving a plateau of $\overline{T} = 1$ year.
	But, starting from $\tau \approx 80$ days, the smoothing reduces significantly the amplitude of the 1-year oscillation, so that the 4-year oscillation starts to prevail, leading to the final plateau at $\overline{T} \approx 4$ years.

	\section{Results}
	
	\subsection{Variation of $\overline{T}$ with $\tau$}
	In this section we apply our analysis technique to real SAT time series.
	We begin by analysing how $\overline{T}$ varies with $\tau$.
	
	\begin{figure*}[tbp] 
		\centering
		{\includegraphics[width=0.45\textwidth]{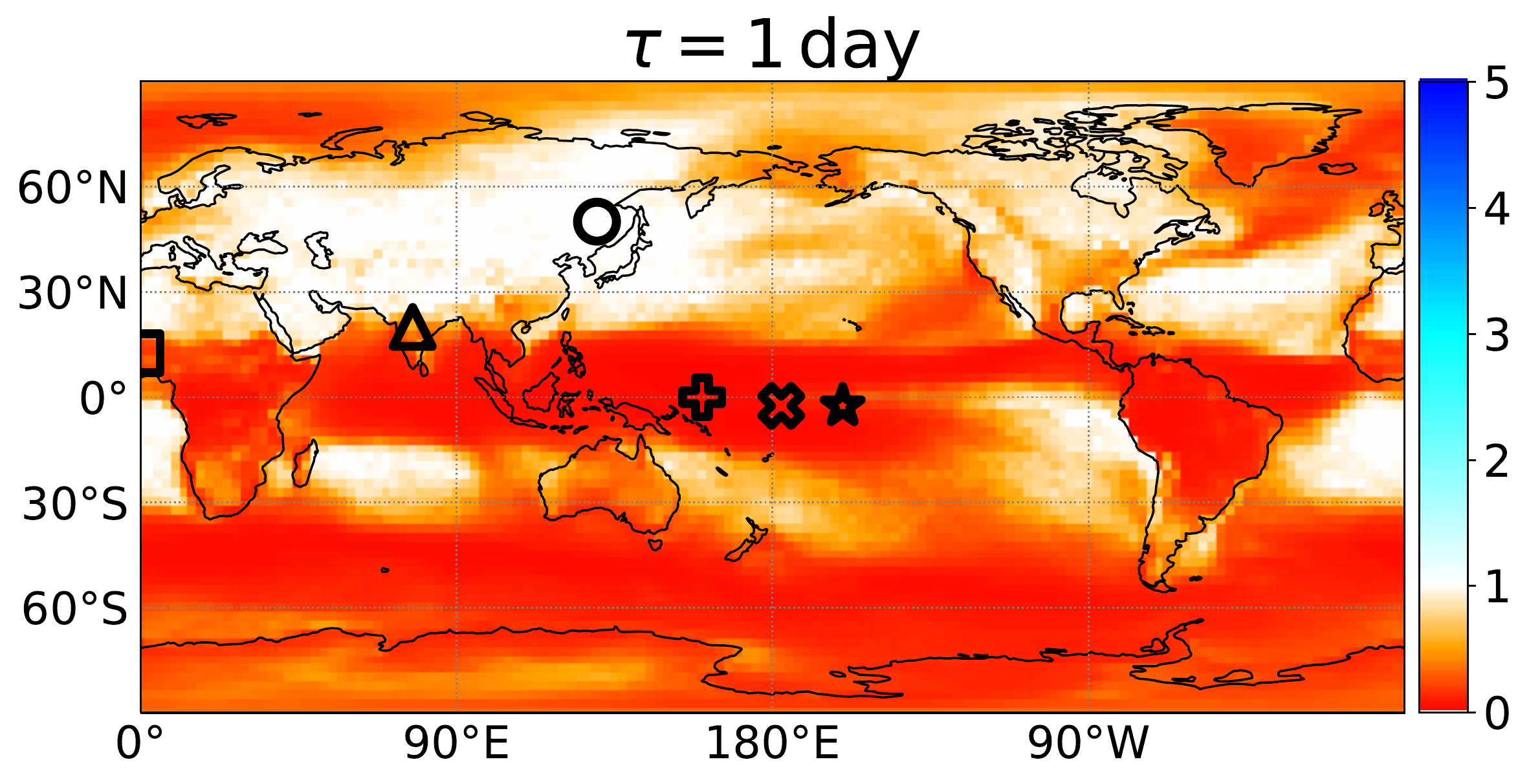}}
		\quad
		{\includegraphics[width=0.45\textwidth]{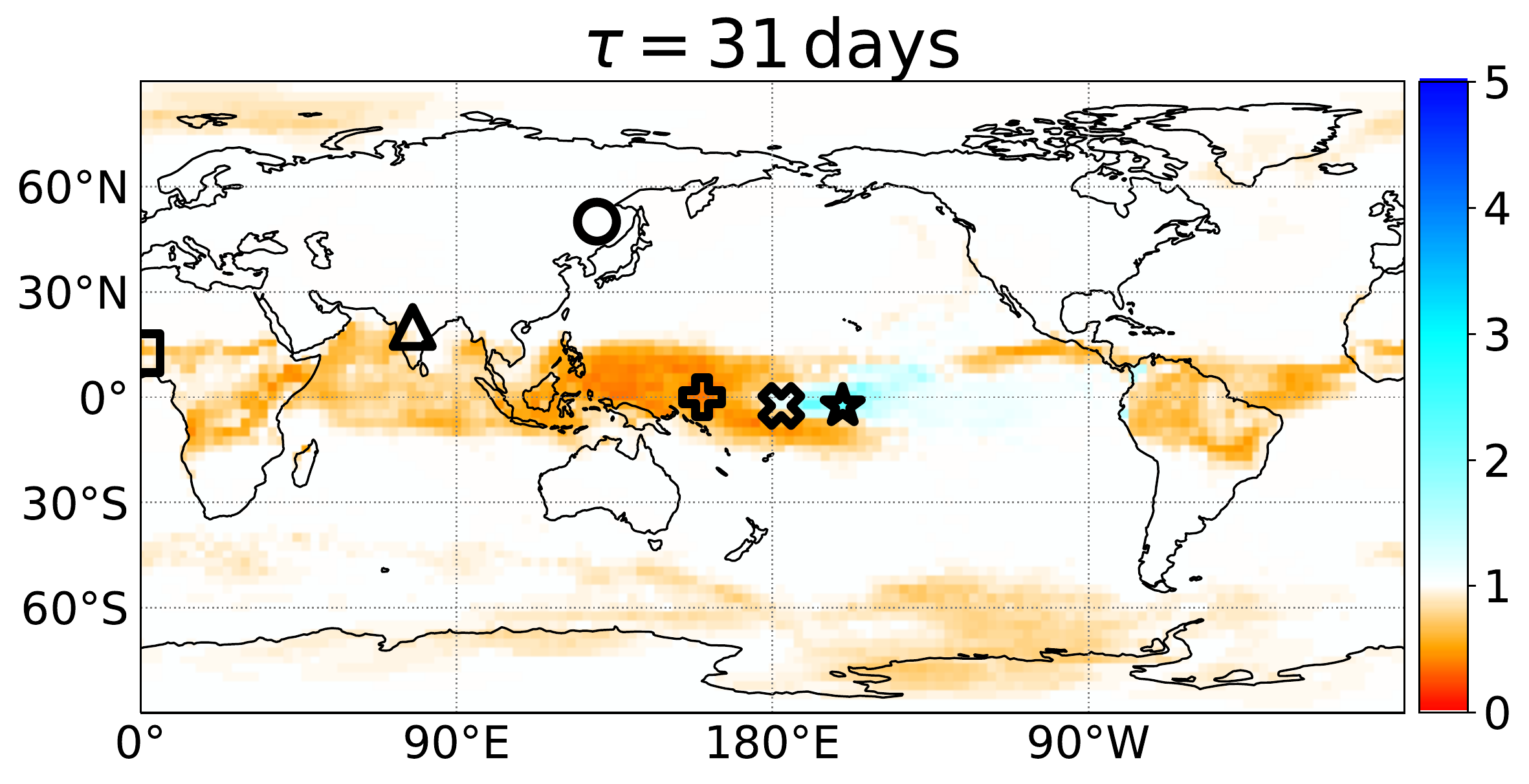}}
		\\
		{\includegraphics[width=0.45\textwidth]{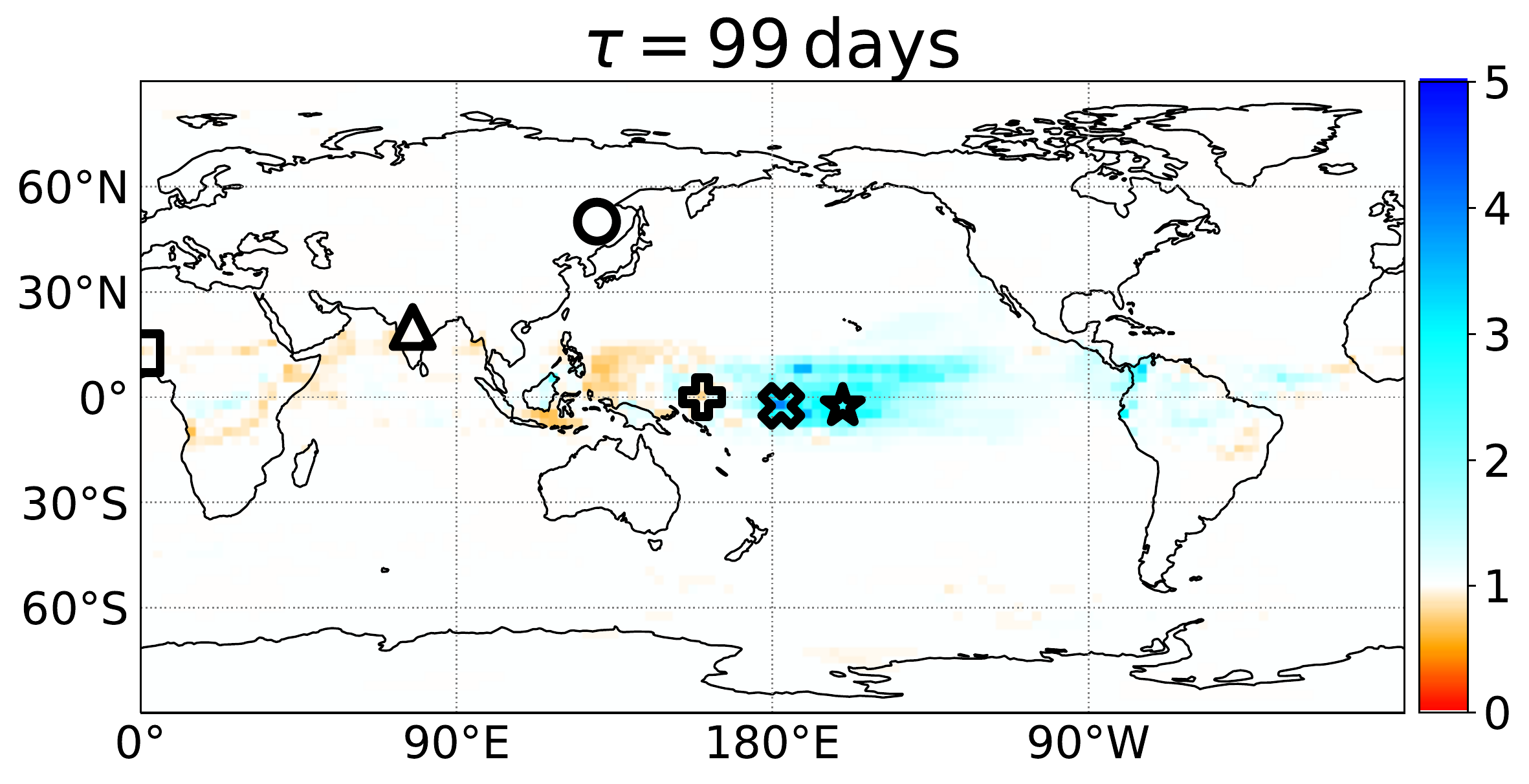}}
		\quad
		{\includegraphics[width=0.45\textwidth]{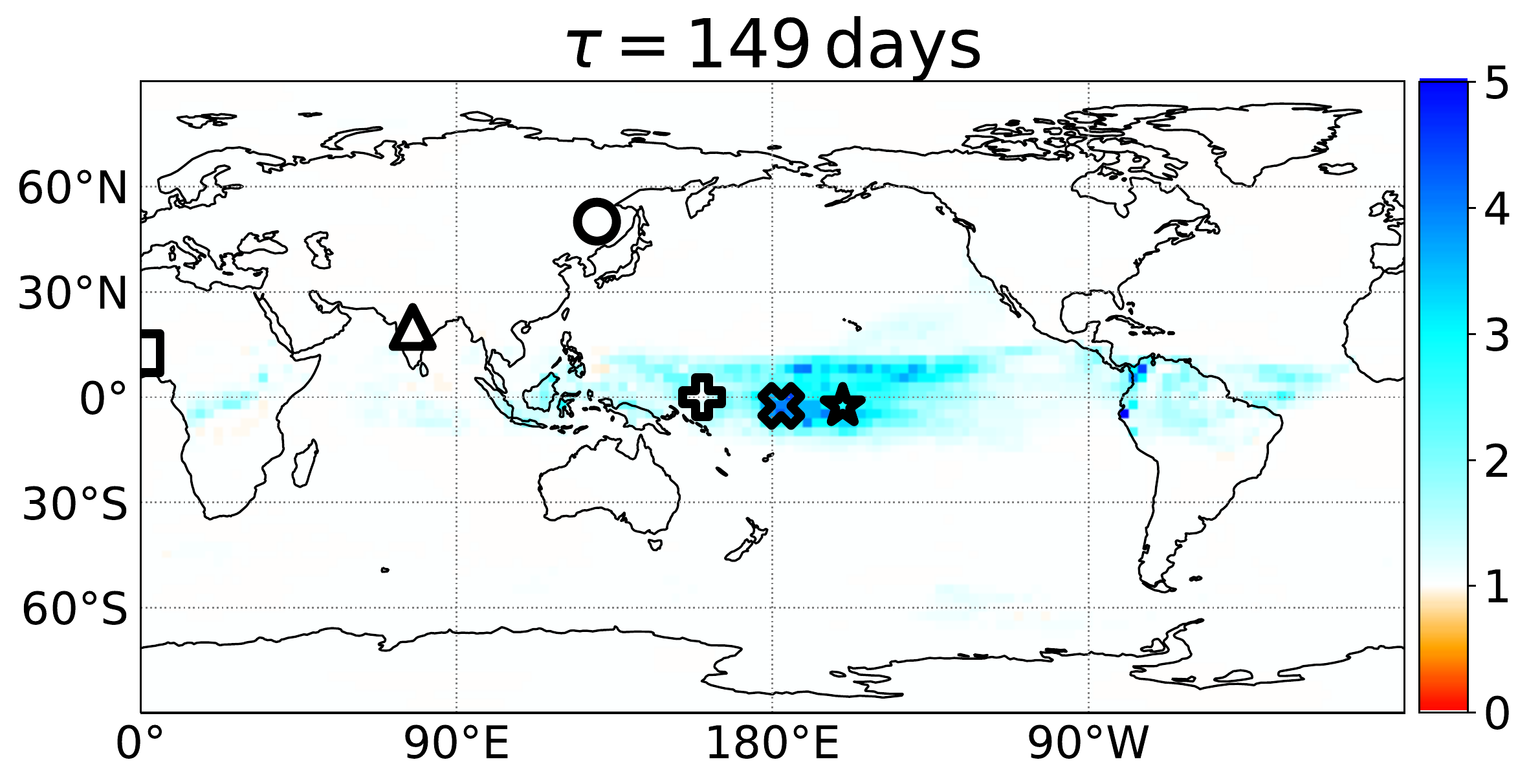}}		
		\caption{Influence of temporal averaging on the mean rotation period.
			The colour maps display $\overline{T}$ (measured in years) computed after smoothing SAT in a moving window of length $\tau=$ 1, 31, 99 and 149 days.
			The symbols indicate the geographical sites discussed in the text: regular (circle), quasi-regular (triangle), double period (square), irregular (plus), El Niño (cross), QBO (star).
			At \url{https://youtu.be/5oX5i5uCm_8} a video shows the evolution of $\overline{T}$ as $\tau$ increases from 1 to 149 days.}
		\label{fig:maps-avgperiod}
	\end{figure*}
	Figure~\ref{fig:maps-avgperiod} displays the colour maps of $\overline{T}$, obtained with $\tau=$ 1, 31, 99 and 149 days. White colour indicates a mean period of 1 year; red colour represents faster dynamics, while blue colour represents slower dynamics. 
	As one could expect, as $\tau$ is increased the regions of fast dynamics are washed out. With $\tau=31$ days, a blue area of slow dynamics begins to emerge in the central Pacific Ocean. This is due to the fact that in this area large-scale variability modes (such as El Niño) produce temperature oscillations with time scale of several years and whose amplitude is larger than the annual oscillation.   
	
	In order to gain insight into how the slow dynamics emerges, we focus the analysis on six geographical sites (indicated with symbols in the colour maps) that are representative of different types of SAT dynamics.
	The symbols in Fig.~\ref{fig:maps-avgperiod} indicate the position of the six sites, which will be referred to as: regular, quasi-regular, double period, irregular, El Niño, and QBO.
	
	\begin{figure*}[tbp] 
		\centering
		\begin{tabular}[t]{rrr}
			{\includegraphics[height=0.25\textwidth]{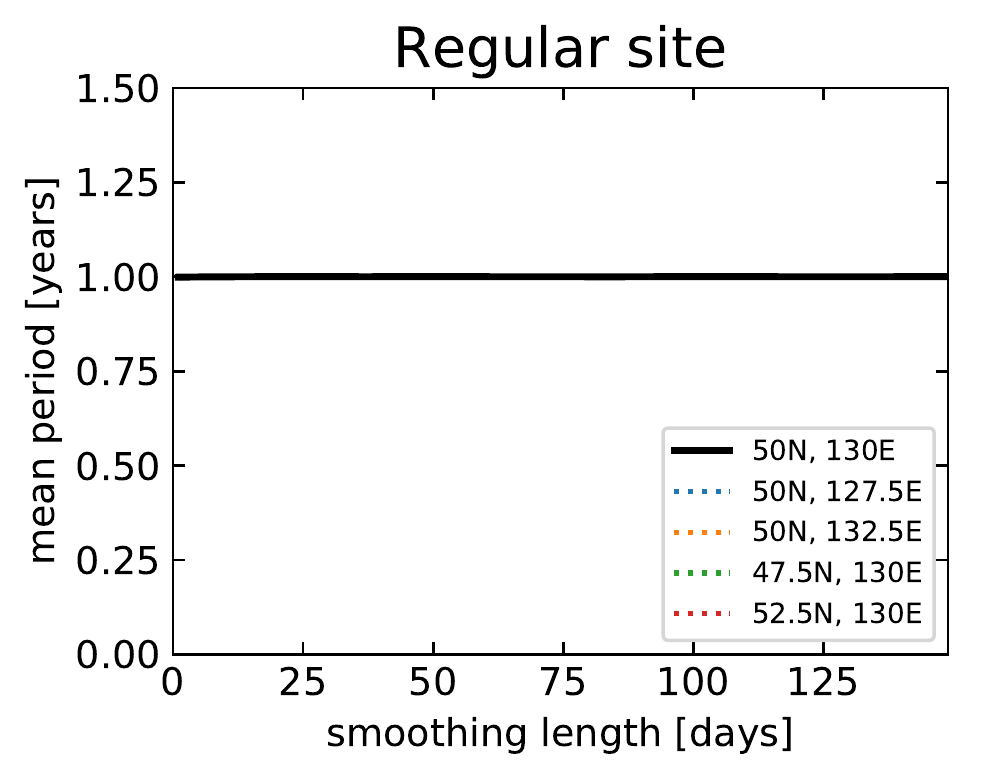}}
			&
			{\includegraphics[height=0.25\textwidth]{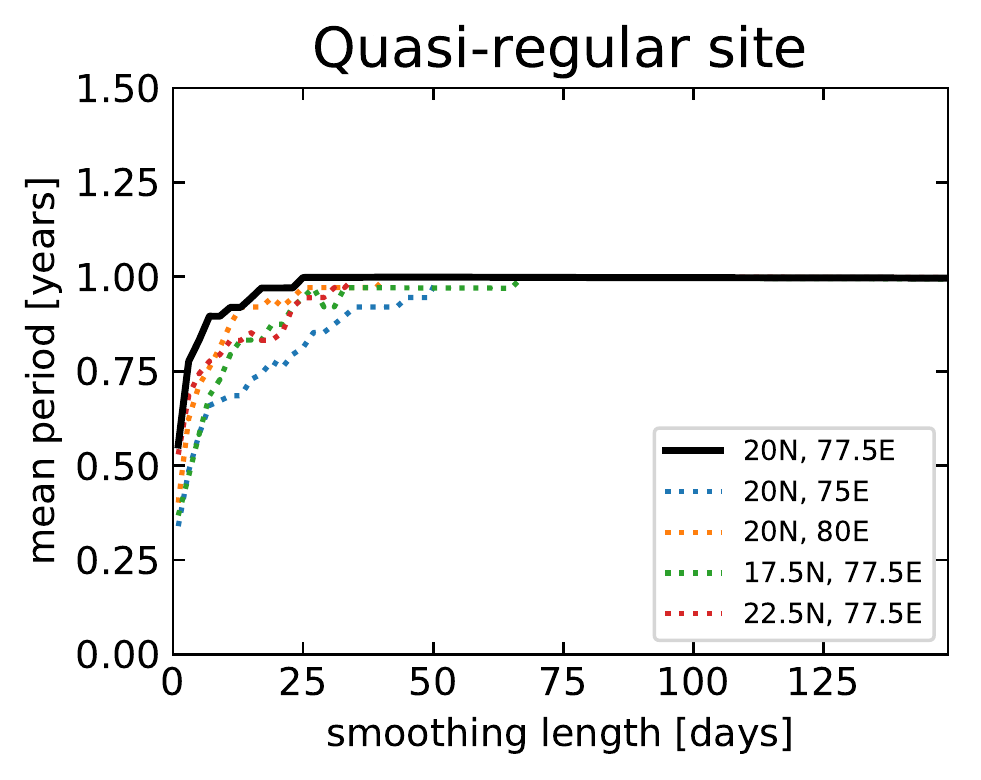}}
			&
			{\includegraphics[height=0.25\textwidth]{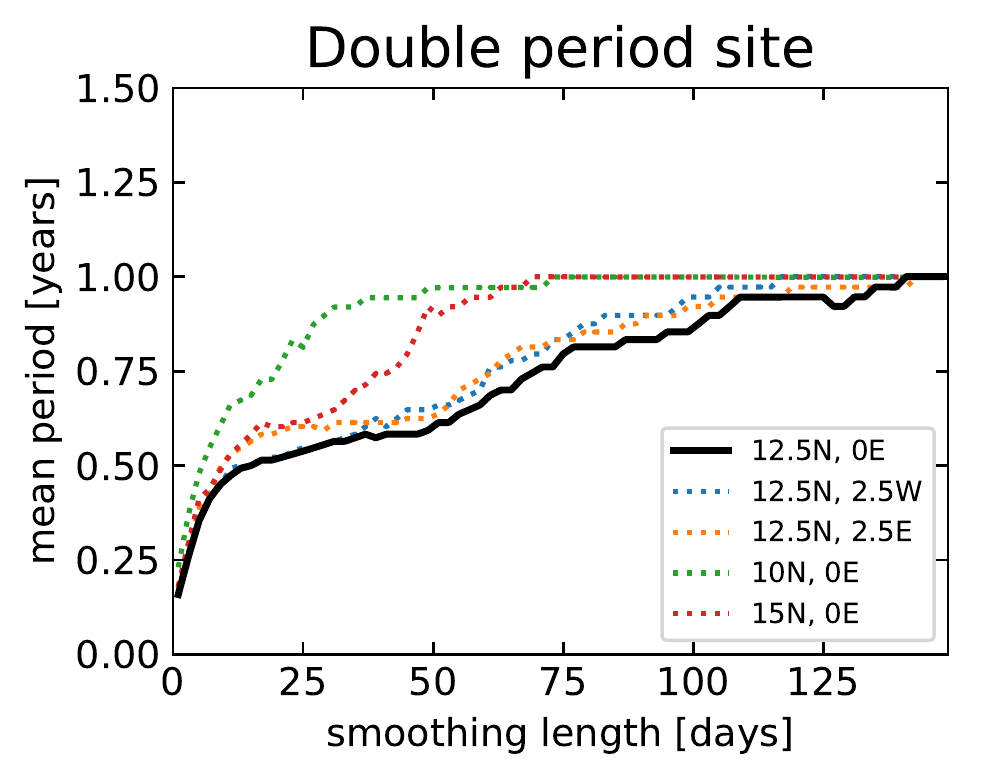}}
			\\
			{\includegraphics[height=0.25\textwidth]{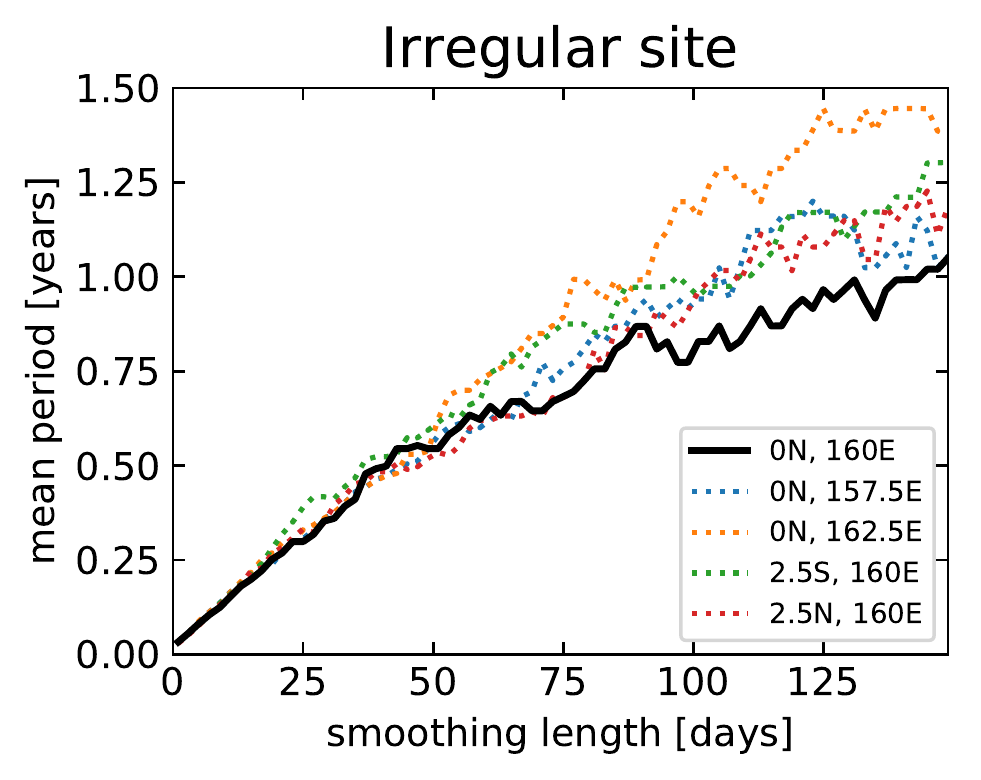}}
			&
			{\includegraphics[height=0.25\textwidth]{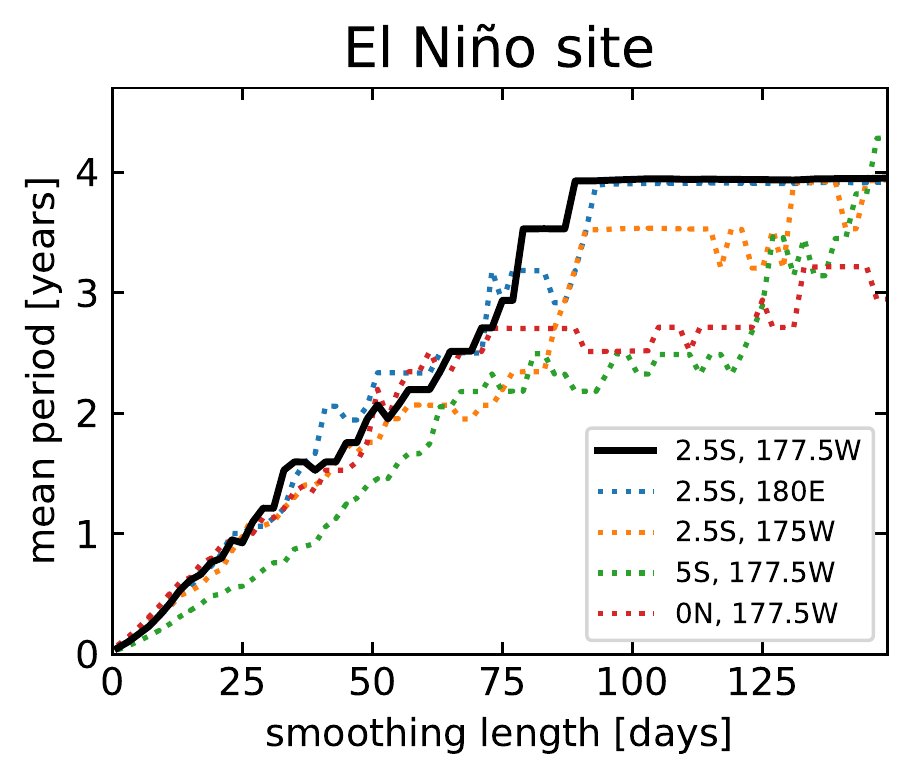}}
			&
			{\includegraphics[height=0.25\textwidth]{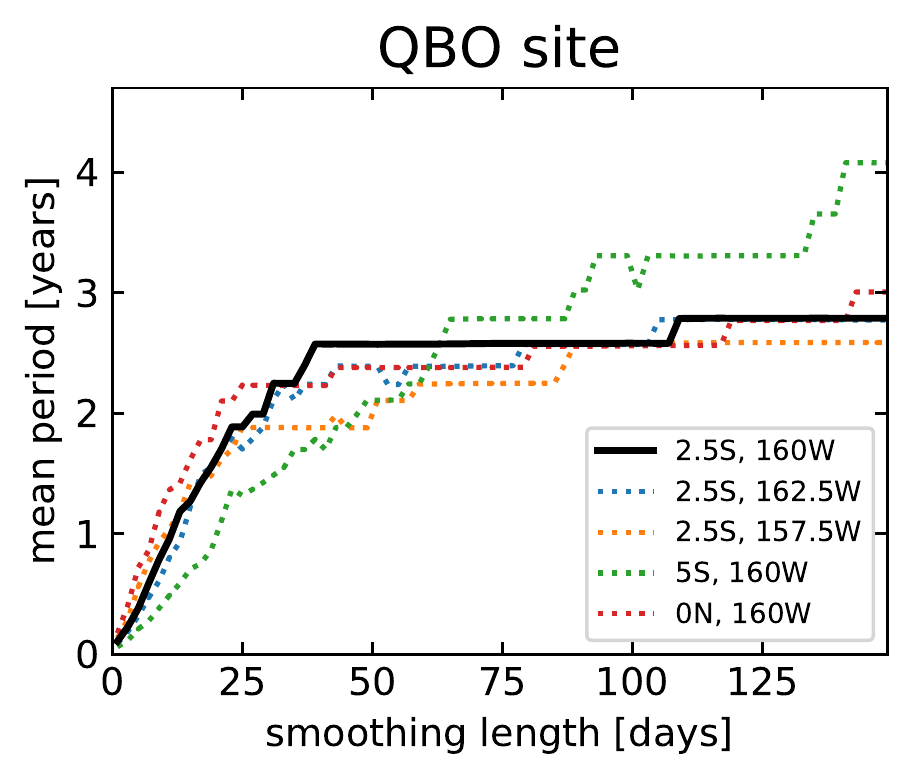}}
		\end{tabular}
		\caption{Variation of the mean period $\overline{T}$ with the length $\tau$ of the smoothing window. The solid line represents the studied site, while the dashed lines represent the four neighbouring sites.}
		\label{fig:periodevol}
	\end{figure*}
	Figure~\ref{fig:periodevol} displays, for the six sites, the variation of $\overline{T}$ with $\tau$.
	We can see that, except for the irregular site, in the other sites $\overline{T}$ shows a \emph{plateau} in some range of values of $\tau$.
	For the regular, quasi-regular and double period sites the  plateau is at $\overline{T}=1$ year.
	The double period site shows also an irregular plateau at $\overline{T} \approx 0.5$ years.
	For the El Niño and QBO sites, the plateau is at $\overline{T} \approx 4$ years and $\overline{T} \approx 2.5$ years, respectively.
	To demonstrate the robustness of these results, in Appendix~\ref{app:compar} we analyse another reanalysis dataset (NCEP Reanalysis 2) and obtain qualitatively similar variation of $\overline{T}$ with $\tau$, with differences that can be attributed to the different spatial resolutions of the two reanalysis.
	As an additional test, we analyse SAT time series with a different temporal resolution: we consider ERA-Interim reanalysis with monthly resolution and find results that are consistent with those obtained from daily SAT data (shown in Fig.~\ref{fig:maps-avgperiod}) with $\tau = 31, 99$ days.

	\subsection{Analysis of Hilbert phase dynamics}
\begin{figure*}[tbp] 
		\centering
		{\includegraphics[width=0.32\textwidth]{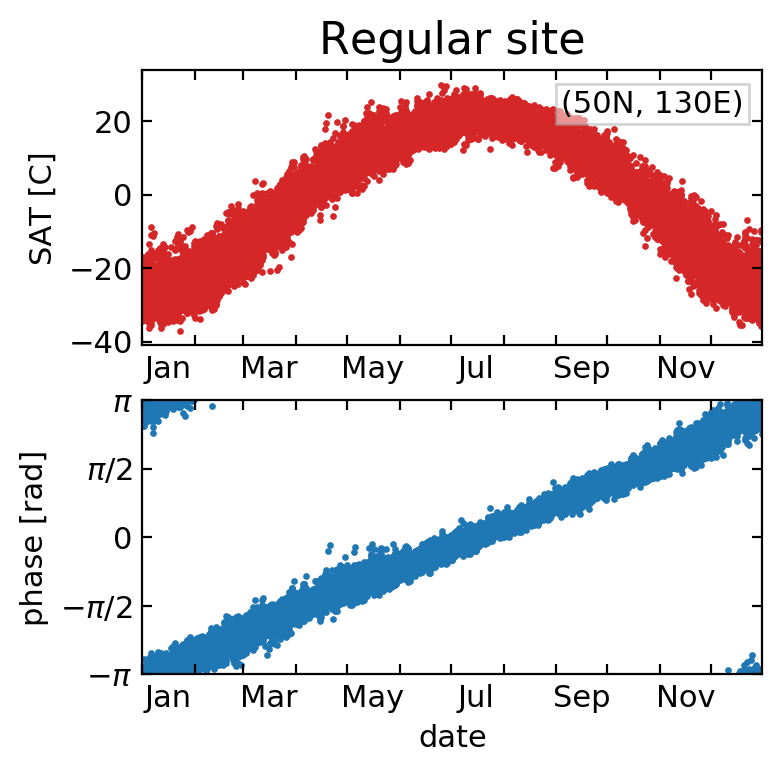}}
		{\includegraphics[width=0.32\textwidth]{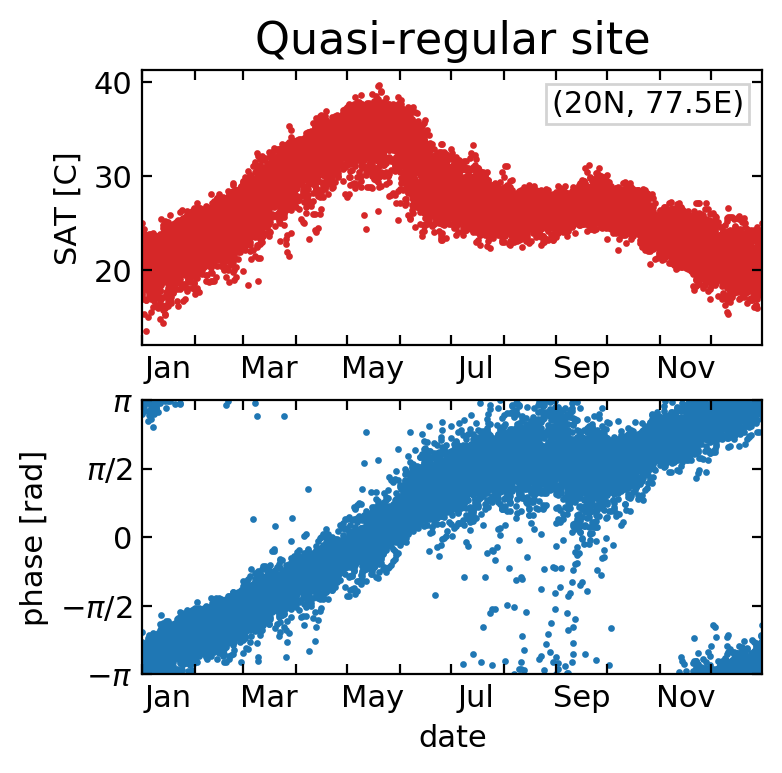}}
		{\includegraphics[width=0.32\textwidth]{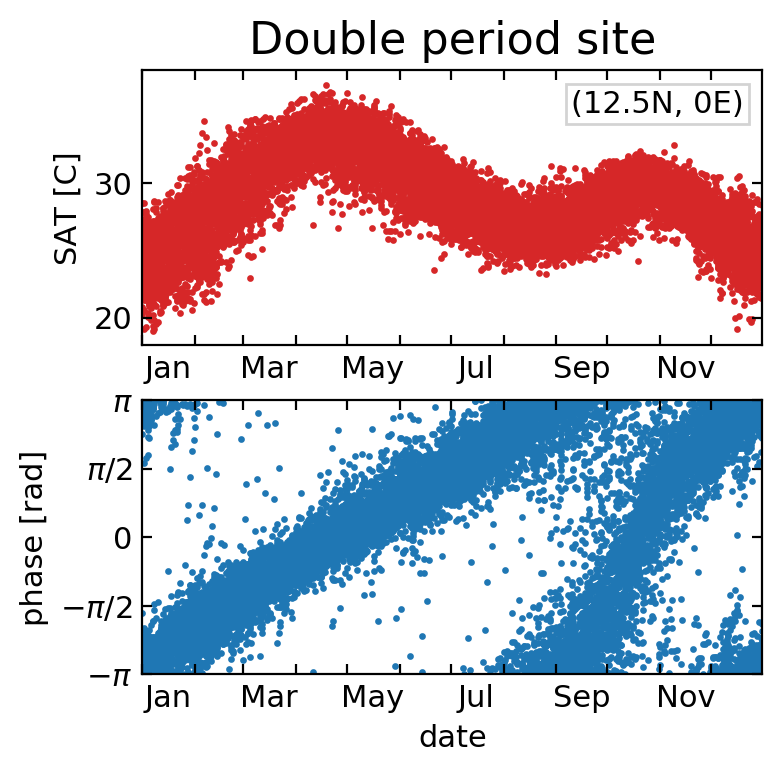}}
		\\
		{\includegraphics[width=0.32\textwidth]{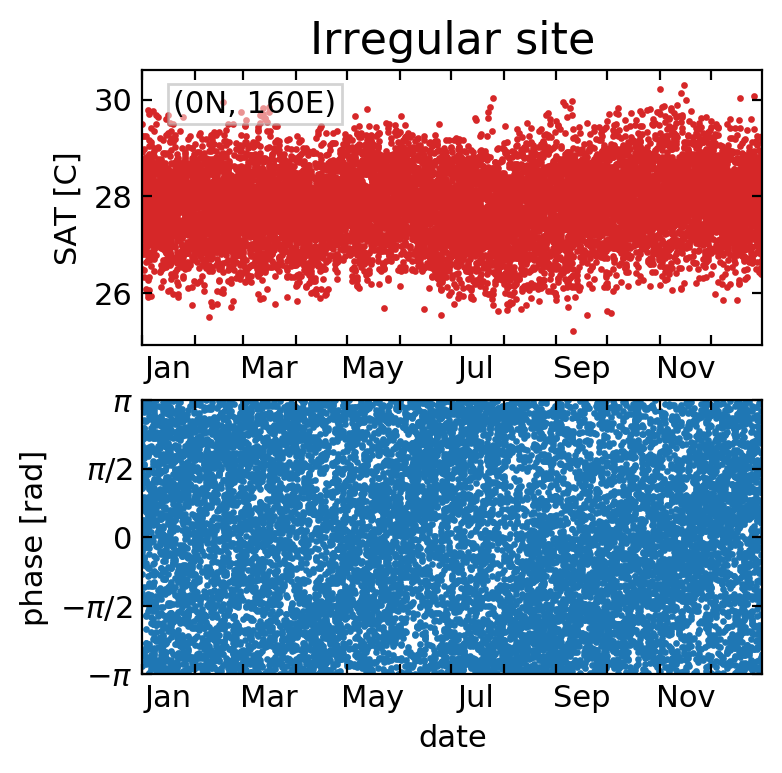}}
		{\includegraphics[width=0.32\textwidth]{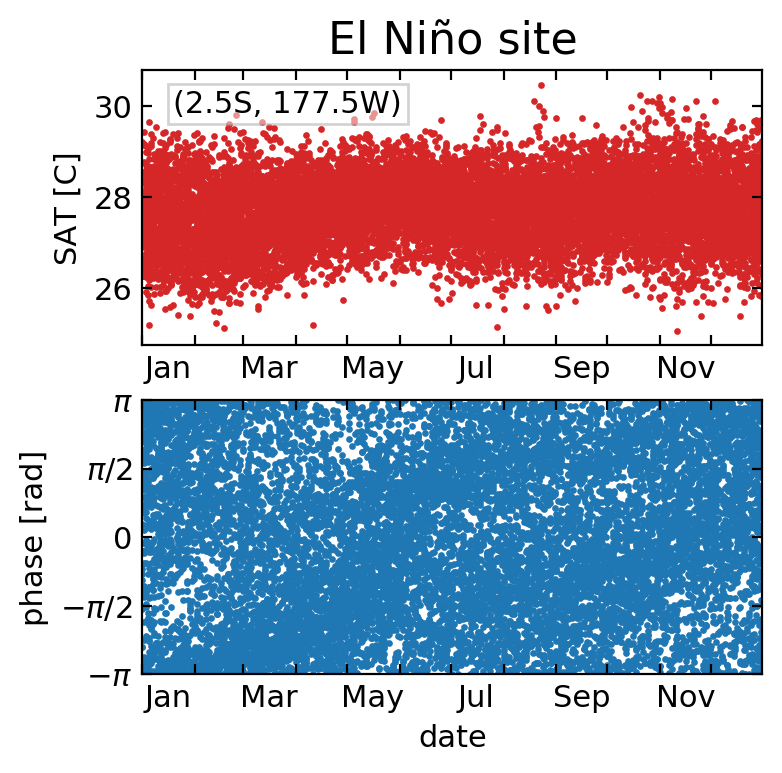}}
		{\includegraphics[width=0.32\textwidth]{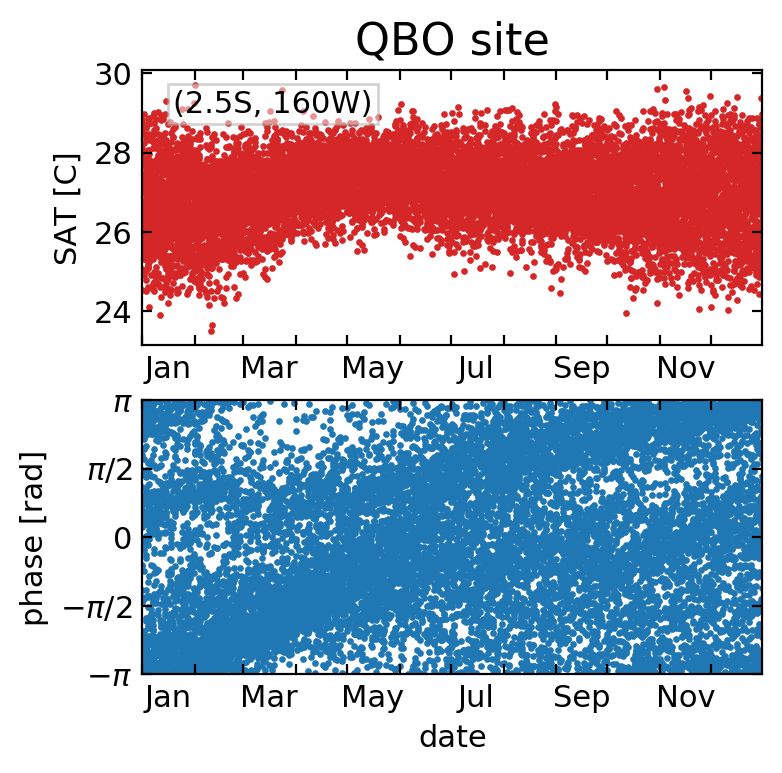}}
		\caption{Hilbert analysis applied to SAT series without pre-filtering.
			In each site, the scatter plots display SAT vs.\ date of the year (upper panels, in red) and Hilbert phase vs.\ date of the year (lower panels, in blue).
			We also indicate the coordinates of each site.}
		\label{fig:map-sat-phase-nosmooth}
	\end{figure*}

\begin{figure*}[tbp] 
		\centering
		\includegraphics[width=0.32\textwidth]{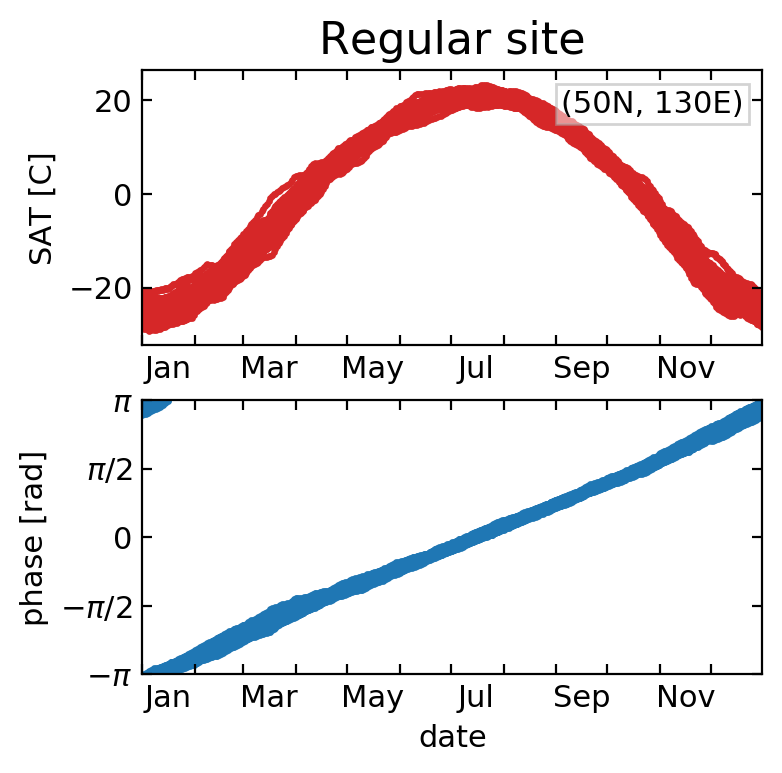}
		\includegraphics[width=0.32\textwidth]{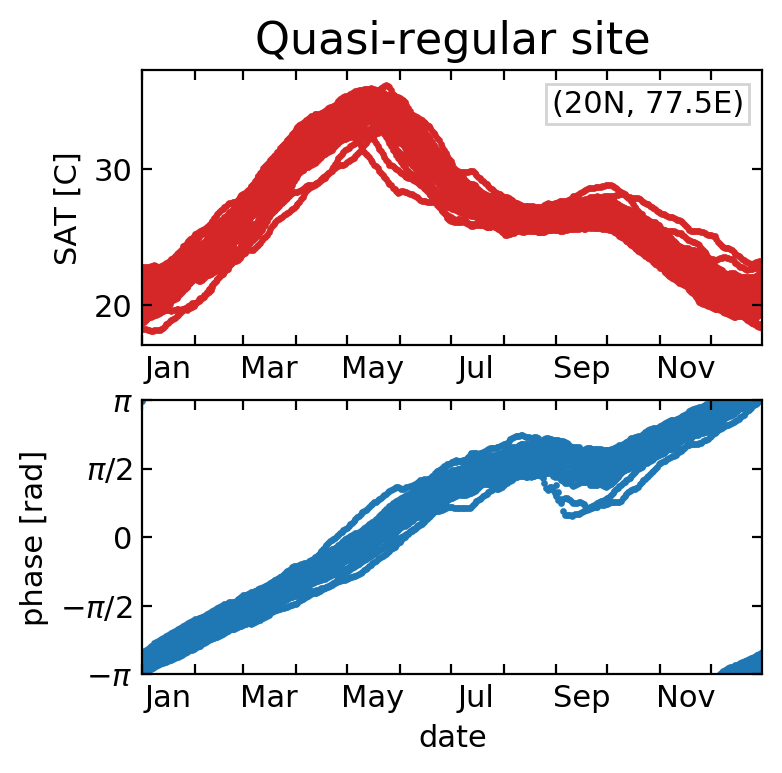}
		\includegraphics[width=0.32\textwidth]{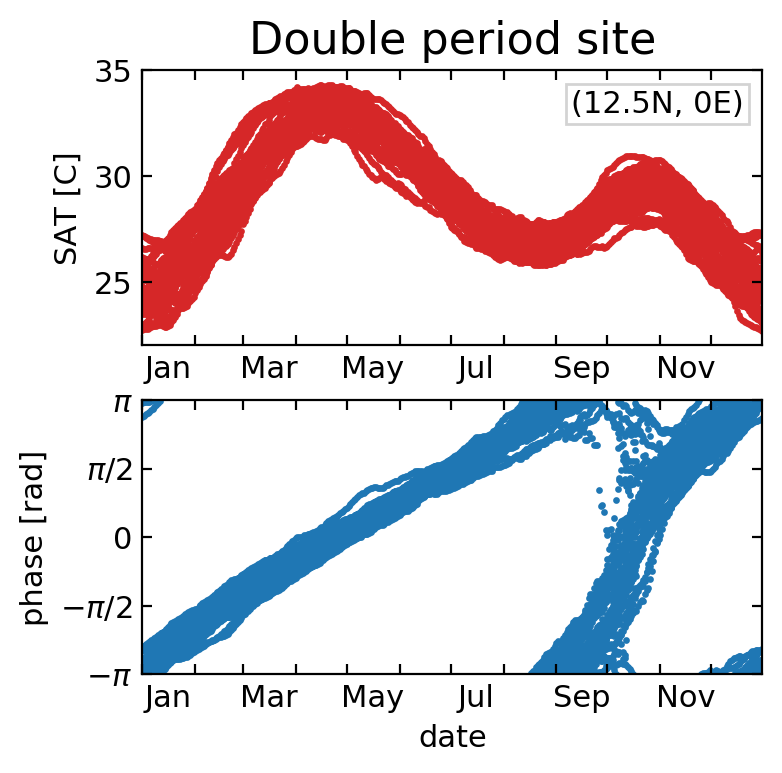}
		\\
		\includegraphics[width=0.32\textwidth]{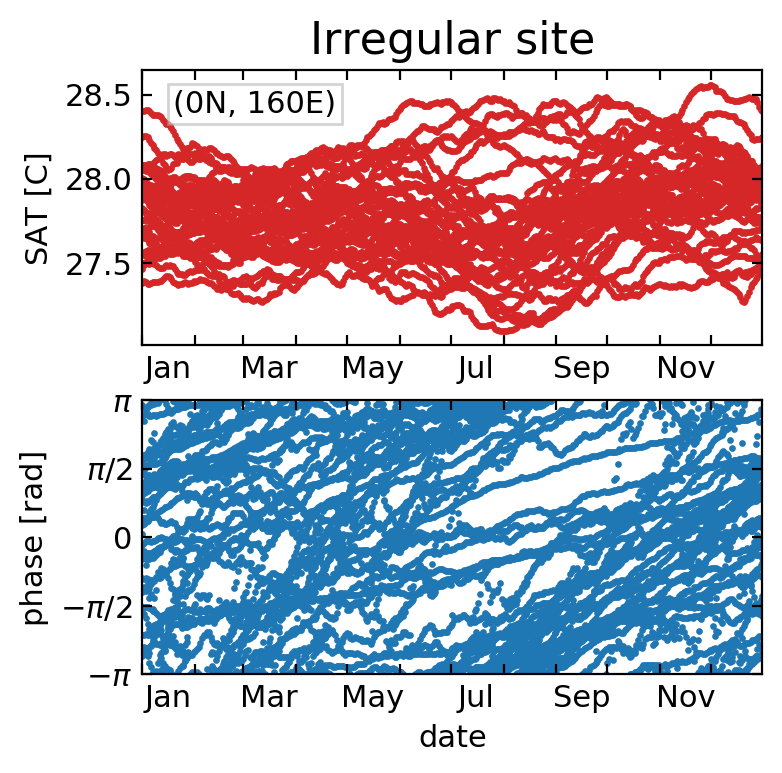}
		\includegraphics[width=0.32\textwidth]{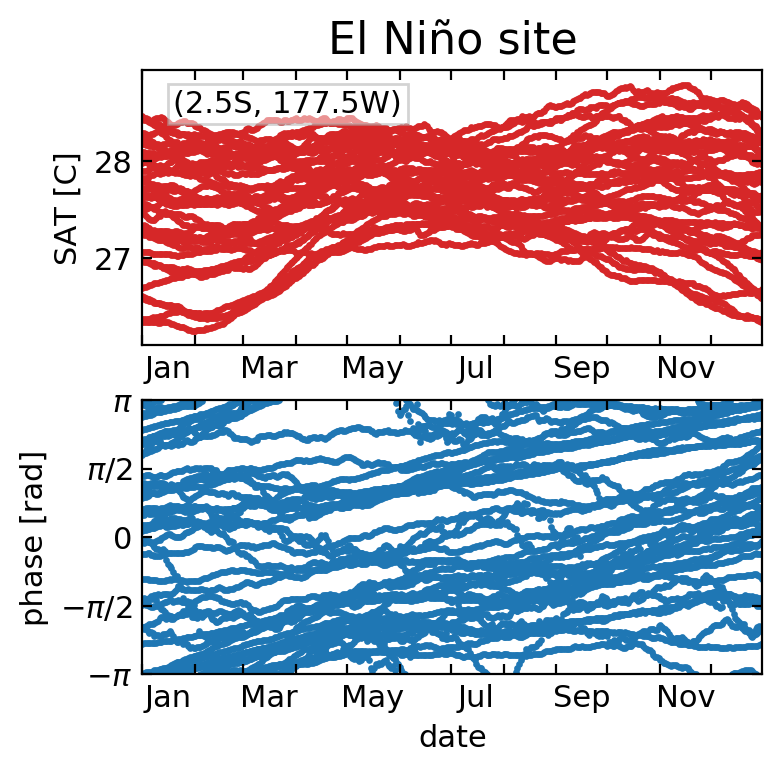}
		\includegraphics[width=0.32\textwidth]{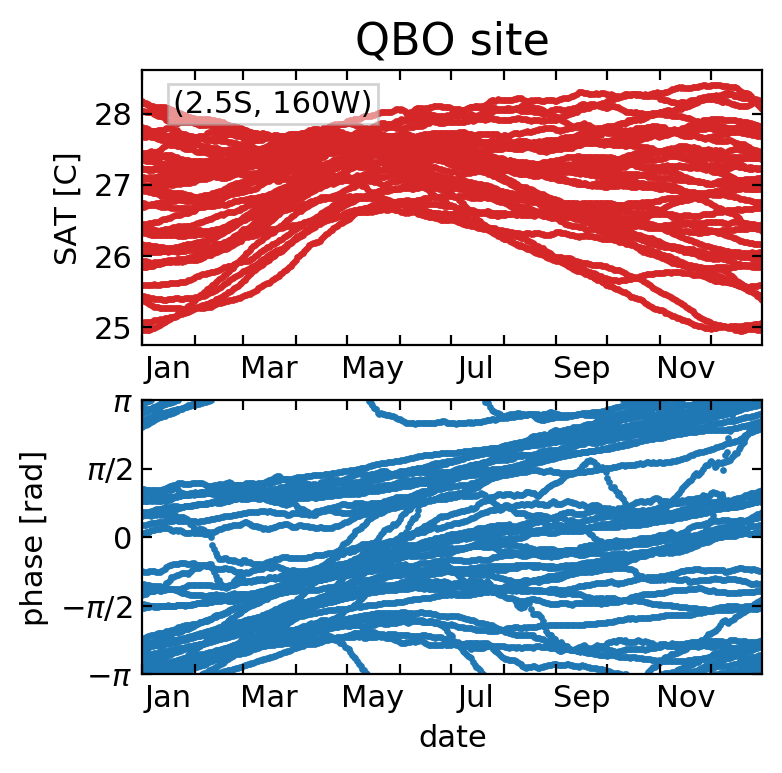}
		\caption{Hilbert analysis applied to pre-filtered SAT series.
			As in Fig.~\ref{fig:map-sat-phase-nosmooth}, but using a smoothing window of length $\tau$ in the plateaus shown in Fig.~\ref{fig:periodevol}: $\tau = 41 \textrm{ days}$ for regular, quasi-regular and double period sites; and $\tau = 101 \textrm{ days}$ for irregular, El Niño and QBO sites.
			We remark that, in the red panels (SAT vs.\ date of the year), the vertical ranges are different from the ones in Fig.~\ref{fig:map-sat-phase-nosmooth}.}
		\label{fig:map-sat-phase-smooth}
	\end{figure*}

	In order to demonstrate that the plateau uncovers underlying regularity in SAT dynamics, we represent for each site the scatter plot of SAT vs.\ date of the year and Hilbert phase vs.\ date of the year, without pre-smoothing (Fig.~\ref{fig:map-sat-phase-nosmooth}) and with pre-smoothing (Fig.~\ref{fig:map-sat-phase-smooth}), using values of $\tau$ that are in the plateau in each site (in the irregular site there is no plateau and thus we use the largest value of $\tau$).
	
	In Fig.~\ref{fig:map-sat-phase-nosmooth} the width of the SAT curves is a measure of the interannual variability in each location (note the different vertical scales of the SAT panels).
	We can observe that the continental extratropical site (the regular one) has larger interannual variability than the two continental tropical sites (quasi-regular and double period), which in turn have larger interannual variability than the three tropical oceanic sites (irregular, El Niño, and QBO). This is consistent with current understanding of atmospheric variability.
	Regarding the seasonal variations, the continental sites show well defined evolutions with one or two maxima that are well captured in the phase scatter plots.
	On the contrary, the tropical oceanic sites show a weak or null seasonal cycle.
	The two easternmost sites show an interannual variability that is maximum during boreal winter and minimum during boreal spring.
	In Fig.~\ref{fig:map-sat-phase-smooth} we see that clear structures emerge in the scatter plots of SAT vs.\ date and phase vs.\ date, when pre-filtering SAT using a window of specific length $\tau$.
	Taken together, Figs.~\ref{fig:periodevol}, \ref{fig:map-sat-phase-nosmooth} and \ref{fig:map-sat-phase-smooth} allow us to uncover and to characterise SAT regularity in each site.
	
	In the regular site, we see in Fig.~\ref{fig:periodevol} that $\overline{T}$ as a function of $\tau$ is remarkably constant. 
	The relation between phase and date of the year, even for the raw SAT (Fig.~\ref{fig:map-sat-phase-nosmooth}), is a clear linear growth that gives one cycle per year, and we see in Fig.~\ref{fig:map-sat-phase-smooth} that taking a temporal average doesn't change this linear behaviour.
	This is due to the fact that in this site the dominating mode of SAT oscillation is produced by solar forcing, which in this region has a period of one year.
	The phase dynamics of the regular site is characteristic of continental extratropical climate, dominated by the annual cycle of solar forcing.
	
	In the quasi-regular site, located in India, we see in Fig.~\ref{fig:periodevol} that without smoothing we get $\overline{T} \approx 0.5\text{ years}$.
	As $\tau$ increases, $\overline{T}$ increases until it reaches a stable plateau of $\overline{T} = 1\text{ year}$ at $\tau = 25\text{ days}$.
	In the phase-date relation obtained without smoothing, we see that the phase has additional cycles during the year: there are years with just one cycle, and years with two or more cycles.
	If the smoothing window is long enough (in Fig.~\ref{fig:map-sat-phase-smooth} we used $\tau = 41\text{ days}$), these additional cycles are washed out, resulting in a phase-date relation with one cycle per year. We interpret these results as due to the fact that SAT in this site has a component with year periodicity, and also half-year periodicity and faster variability, which are reduced by the 41-day filtering that leaves the time series dominated by the annual cycle.
	We note in Fig.~\ref{fig:map-sat-phase-smooth} that the phase increase is not linear during June-September, a result that captures the effect of the Indian monsoon that produces the small plateau in temperature in the same months.
	
	The double period site is located in another monsoon region, in this case the West African monsoon.
	The plot of the mean period starts, for no smoothing, with $\overline{T} \approx 0.2\text{ years}$. As $\tau$ increases, $\overline{T}$ has a steep increase until $\tau \approx 10\text{ days}$, then it shows a slower growth for $\tau$ in the range of 10-45 days, with a mean period of $\overline{T} \approx 0.5\text{ years}$.
	When $\tau$ is increased further, $\overline{T}$  continues to grow reaching the one year period at $\tau \approx 140\text{ days}$.
	We interpret this behaviour as due to the presence of a component of half-year period whose amplitude is larger than the component with one year periodicity.
	In the SAT-date and phase-date scatter plots, with no smoothing (Fig.~\ref{fig:map-sat-phase-nosmooth}) we see a noisy double cycle, while with 41-day smoothing (Fig.~\ref{fig:map-sat-phase-smooth}) we are left with a dominant half-year oscillatory component.
	If we increase $\tau$, we gradually eliminate the half-yearly component and extract a one year periodicity (not shown).
	The stochastic half year period in this region captures the northward movement of the Intertropical Convergence Zone during boreal summer, which strongly affects the surface temperature leading to peaks before and after the monsoon.
	We need to point out that the quasi-regular and double period sites have similar characteristics.
	We can see (Figs. \ref{fig:map-sat-phase-nosmooth} and \ref{fig:map-sat-phase-smooth}) that their SAT time series are similar and the main difference is whether the second local minima is smaller or larger than the mean of the time series.
	For this reason, there can be variations from one year to the other: the quasi-regular site can have years with a double oscillation and the double period site can have years with only one oscillation.
	It is not a surprising behaviour, since both sites are located in monsoon regions and have a climate with strong interannual variability (influenced, among other factors, by El Niño).
	
	The irregular site is the westernmost of the three sites that we selected in the Pacific Ocean.
	We see that, for no smoothing, $\overline{T}$ starts from a low value ($\overline{T} < 0.1\text{ years}$) and as $\tau$ is increased, $\overline{T}$ increases without revealing any particular time scale, i.e., it does not reach any plateau.
	The phase-date relation with no smoothing does not reveal any structure in the temperature dynamics, although there is a hint of a double period seen as a relative increase in the density of points, as would be expected for a region on the equator.
	If we smooth with $\tau = 101 \text{ days}$, we still get a phase-date plot that doesn't suggest any clear time scale.
	These results indicate that in this region there is no dominating component of any particular period, i.e., SAT time series is consistent with the sum of stochastic processes with different time scales and similar amplitudes.
	
	Next, we analyse the El Niño site, the central of the three sites in Pacific Ocean, located in the so-called cold tongue region. As in the previous site, without smoothing we find that $\overline{T} < 0.1 \text{ years}$ and then $\overline{T}$ increases linearly with $\tau$.
	However, in contrast to the irregular site, here $\overline{T}$ reaches a stable plateau of $\overline{T} \approx 4 \text{ years}$ at $\tau \approx 90 \text{ days}$.
	This means that, if we average SAT over a sufficiently long temporal window, we are left with a dominant oscillation whose period is approximately 4 years.
	In the phase-date relation a 4-year cycle would be represented by a line that cycles 4 times in the horizontal direction while covering vertically the 2$\pi$ phase range. In the phase-date relation calculated with $\tau = 101 \text{ days}$ (Fig.~\ref{fig:map-sat-phase-smooth}) we see a hint of such ordered phase dynamics (in the form of tilted regions with higher density of points), which is not seen in the phase-date relation computed from the raw data (Fig.~\ref{fig:map-sat-phase-nosmooth}).
	This is interpreted as an effect of El Niño, which has a period between 3 to 7 years, with maximum amplitude in the equatorial Pacific cold tongue region.
	
	In the QBO site (the easternmost of the three Pacific sites) $\overline{T}$ starts with a steep increase from $\overline{T} \approx 0.1 \text{ years}$ to $\overline{T} \approx 2.5 \text{ years}$ for $\tau \approx 40 \text{ days}$ where there is a plateau.
	We interpret here that Hilbert phase analysis captures the effects of the QBO oscillation, which consists in the alternation of zonal winds between easterlies and westerlies in the tropical stratosphere, with a mean period of 28-29 months, and which has been shown to  influence SAT~\cite{bib:QBO1, bib:QBO3}.
	In the phase-date relation calculated after averaging SAT in a window of $\tau=101$ days (Fig.~\ref{fig:map-sat-phase-smooth}) we see two tilted bands with higher density of points (consistent with a noisy cycle with $2.5$ year period), which are not so evident in the phase-date relation obtained from the raw SAT (Fig.~\ref{fig:map-sat-phase-nosmooth}).

	\subsection{Fourier analysis}
	
	\begin{figure*}[tbp] 
		\centering
		\begin{tabular}[t]{rrr}
			{\includegraphics[height=0.25\textwidth]{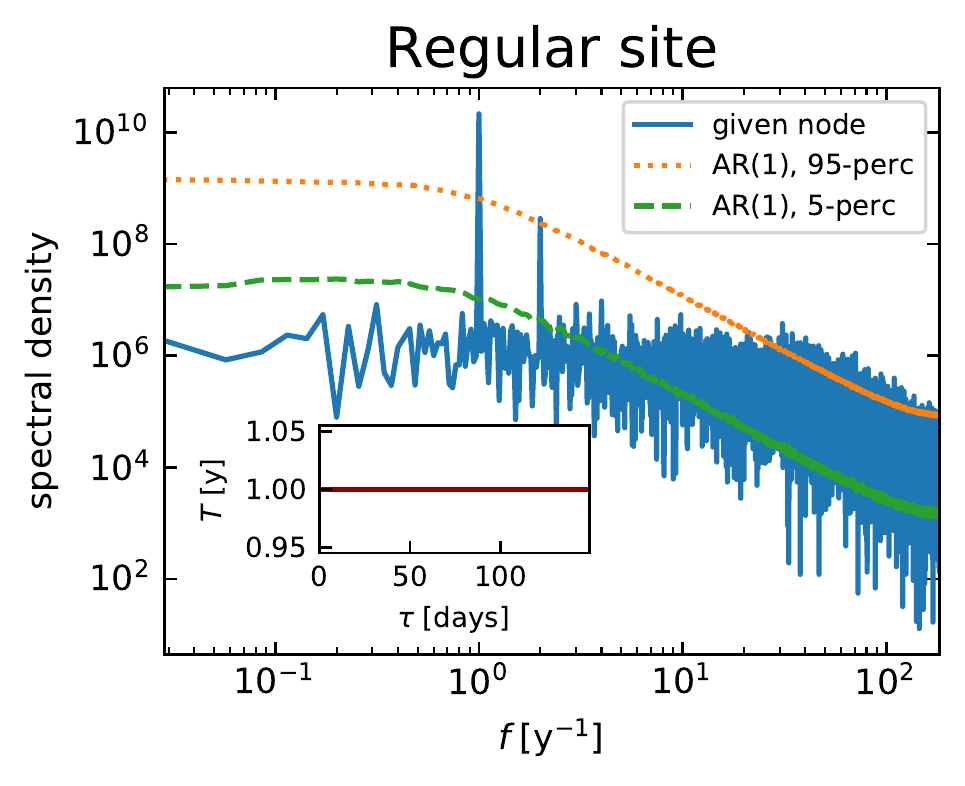}}
			&
			{\includegraphics[height=0.25\textwidth]{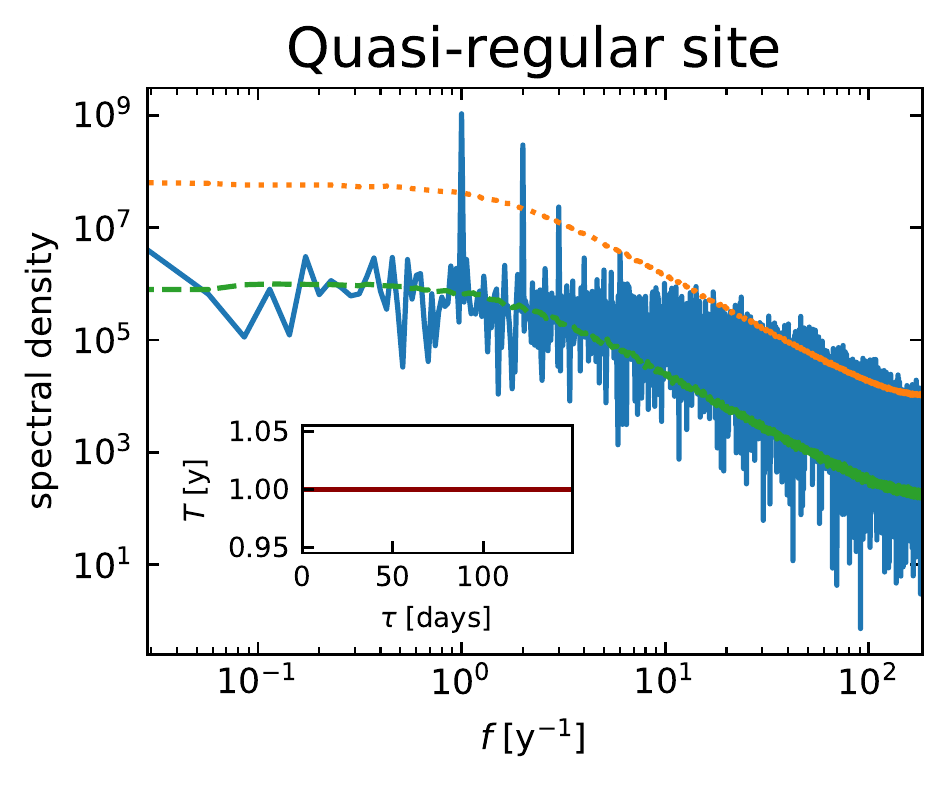}}
			&
			{\includegraphics[height=0.25\textwidth]{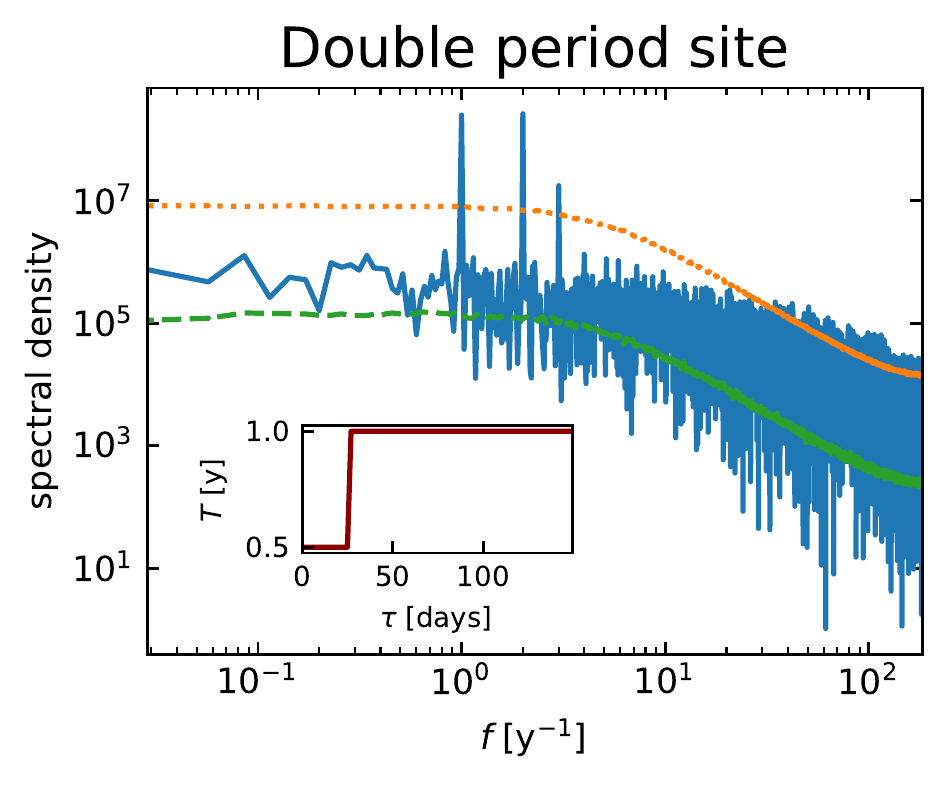}}
			\\
			{\includegraphics[height=0.25\textwidth]{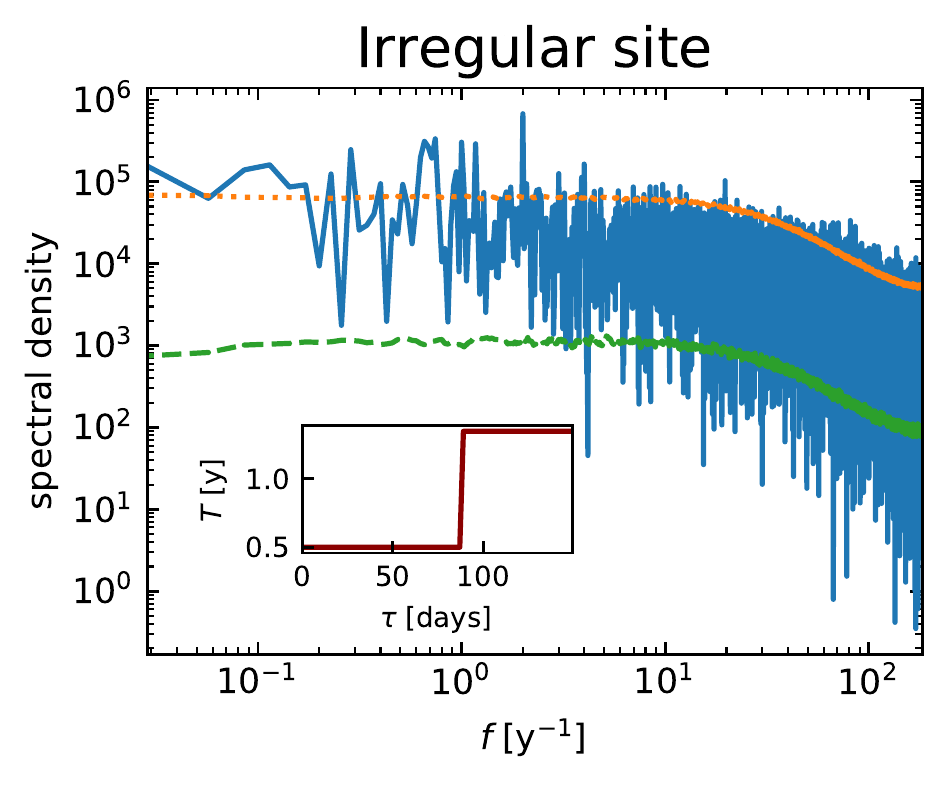}}
			&
			{\includegraphics[height=0.25\textwidth]{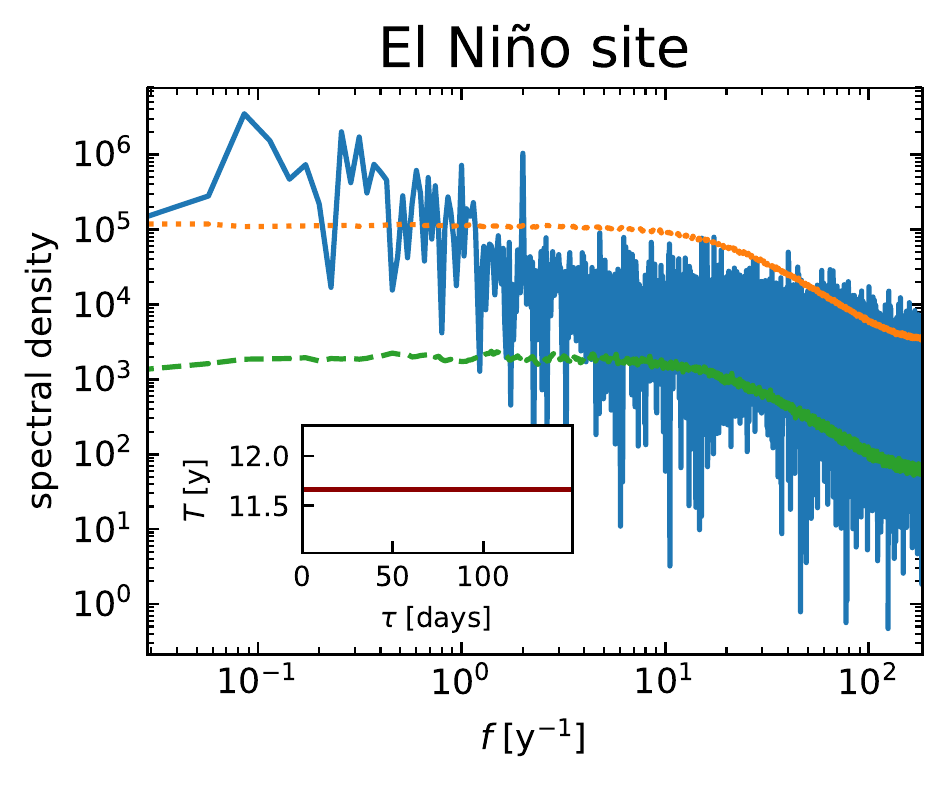}}
			&
			{\includegraphics[height=0.25\textwidth]{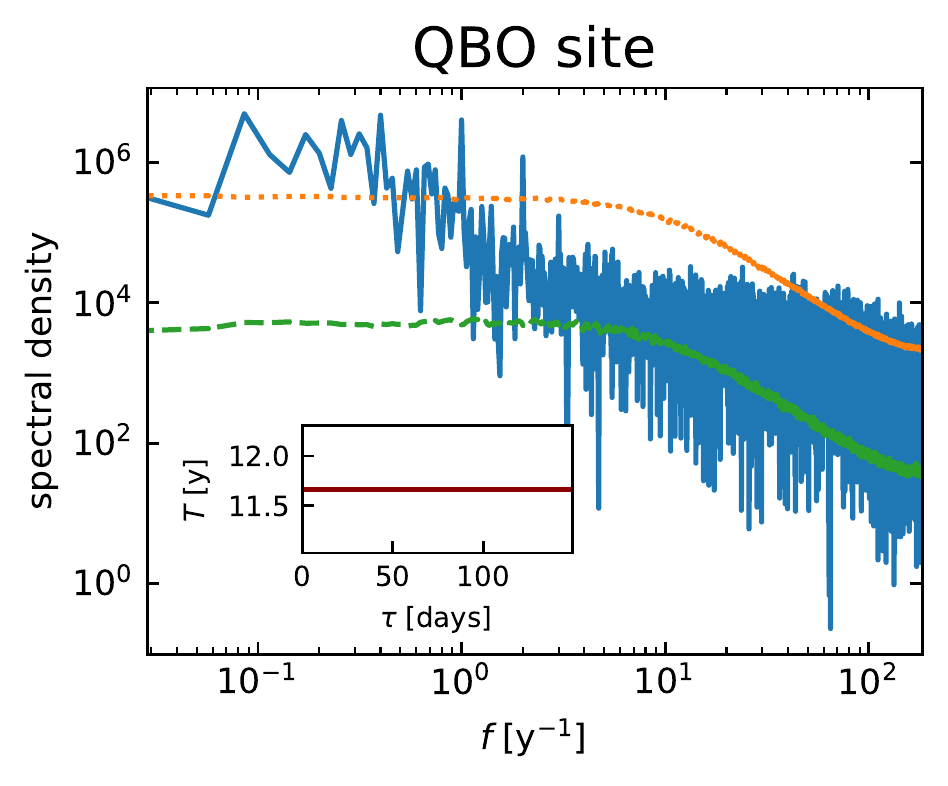}}
		\end{tabular}
		\caption{{Fourier analysis of SAT time series.}
			The panels display the power spectrum of the SAT time series in the different sites (without pre-smoothing, in blue).
			We test the statistical significance of the Fourier peaks with 1000 realisations of AR(1) series with the same variance and autocorrelation (at lag 1) as the SAT series.
			We calculate the spectra of the 1000 realisations and then take the 5 and 95 percentiles.
			We also smooth these two resulting spectra with a moving average of length 5.
			The green and orange lines indicate the 5 and 95 percentiles, respectively, obtained after these calculations.
			We consider a peak of the original spectrum (in blue) to be statistically significant if it is higher than the 95-percentile spectrum (in orange).
			We analyse the influence of smoothing by computing the spectra after averaging the SAT time series in a window of length $\tau$, and calculate the dominant period, $T$, as the inverse of the frequency of the highest significant peak in the spectrum.
			The insets show the variation of $T$ with $\tau$. We note that the gradual emergence of temporal regularity that occurs as SAT variability is washed out is not detected in this way, because $T$ is either constant or it changes abruptly with $\tau$.}
		\label{fig:Fourier}
	\end{figure*}
	In this section we investigate if Fourier spectral analysis can detect the underlying temporal regularities detected through Hilbert phase analysis.
	To do this, we analyse how the highest peak, $f$, in the power spectrum depends on $\tau$.
	The plot of $T = 1 / f$ vs.\ $\tau$ presented in Fig.~\ref{fig:Fourier} reveals the same variation found with Hilbert analysis \emph{only in the regular site}; in the other sites, the gradual variation of $T$ with $\tau$ and the plateaus found are not detected.
	
	In the regular and in the quasi-regular sites Fourier gives a period equal to 1 year, regardless of $\tau$, while we have seen in Fig.~\ref{fig:periodevol} that the quasi-regular site starts with a faster dynamics and then the period rapidly increases to the stable value of 1 year. In the double period site, the period starts at $T = 0.5\text{ years}$ and remains constant as $\tau$ increases until $\tau = 25\text{ days}$, where it suddenly jumps to $T = 1\text{ year}$. Thus, here we lose the gradual variation captured with Hilbert analysis. In the irregular site, we have a similar sudden jump, from $T = 0.5\text{ years}$ to $T \approx 1.3\text{ years}$ at $\tau \approx 90\text{ days}$. In contrast, Hilbert analysis and the phase-date relation tell us that there is no dominating oscillatory component with any smoothing length. In El Niño and QBO sites, $T \approx 11.7\text{ years}$, regardless of $\tau$, which is due to the fact that the Pacific has variability from interannual to interdecadal time scales, and thus, SAT Fourier spectrum has high energy at low frequencies.
	
	We need to say that we are comparing the results obtained from our Hilbert approach with the results obtained from a simplistic application of Fourier transform.
	It could be interesting (but out of the scope of this paper) to use for this comparison more information obtained from Fourier transform (for example, by applying some of the techniques described in \cite{bib:Ghil:adv}).

	\subsection{Classification of SAT dynamics}
	
	\begin{figure}[tbp] 
		\centering
		\includegraphics[width=\columnwidth]{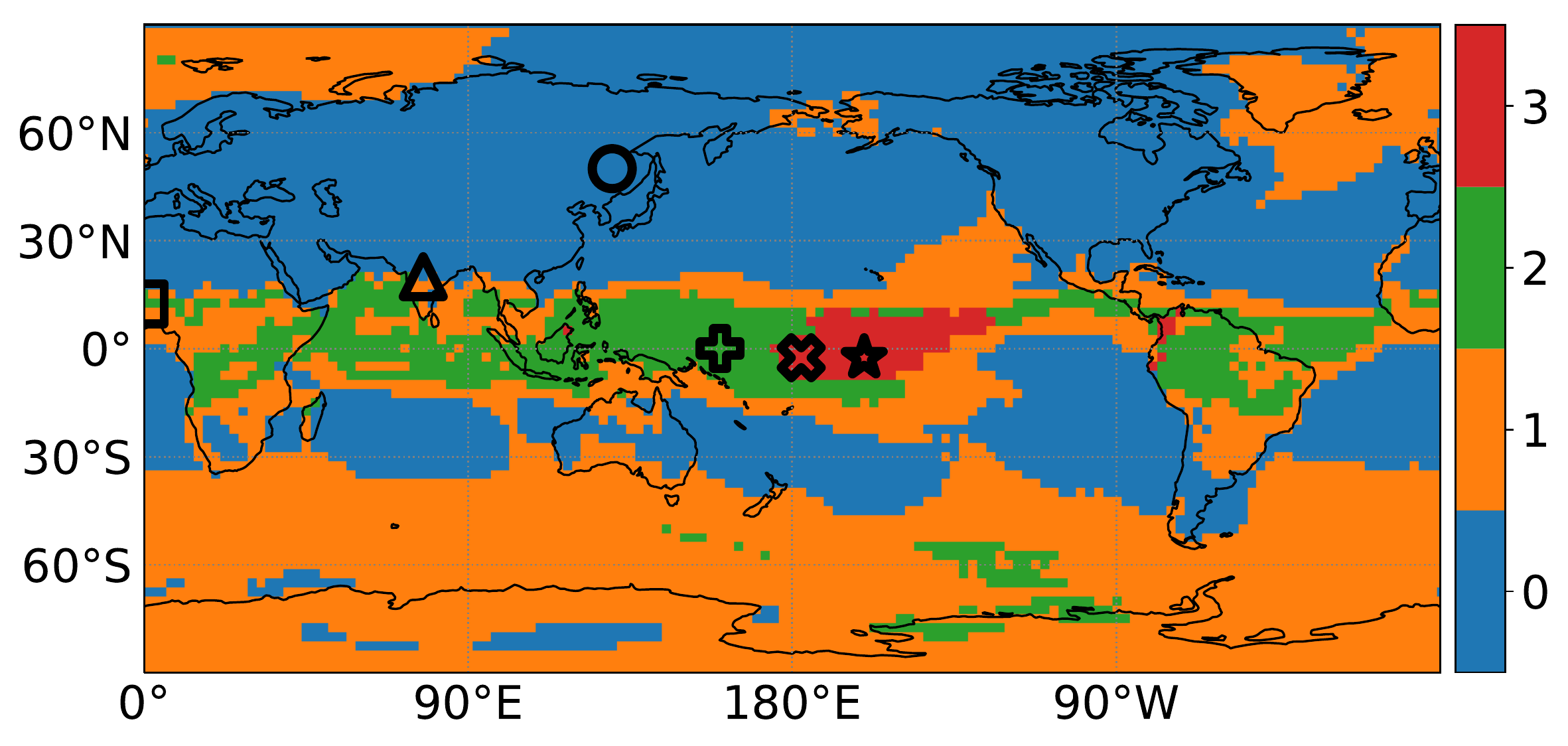}
		\caption{{Classification of the geographical sites in four clusters.}
			The k-means algorithm is used to classify the geographical sites according to their values of $\overline{T}$ for different choices of $\tau$.}
		\label{fig:map:4clusters}
	\end{figure}
	\begin{figure}[tbp] 
		\includegraphics[width=0.49\columnwidth]{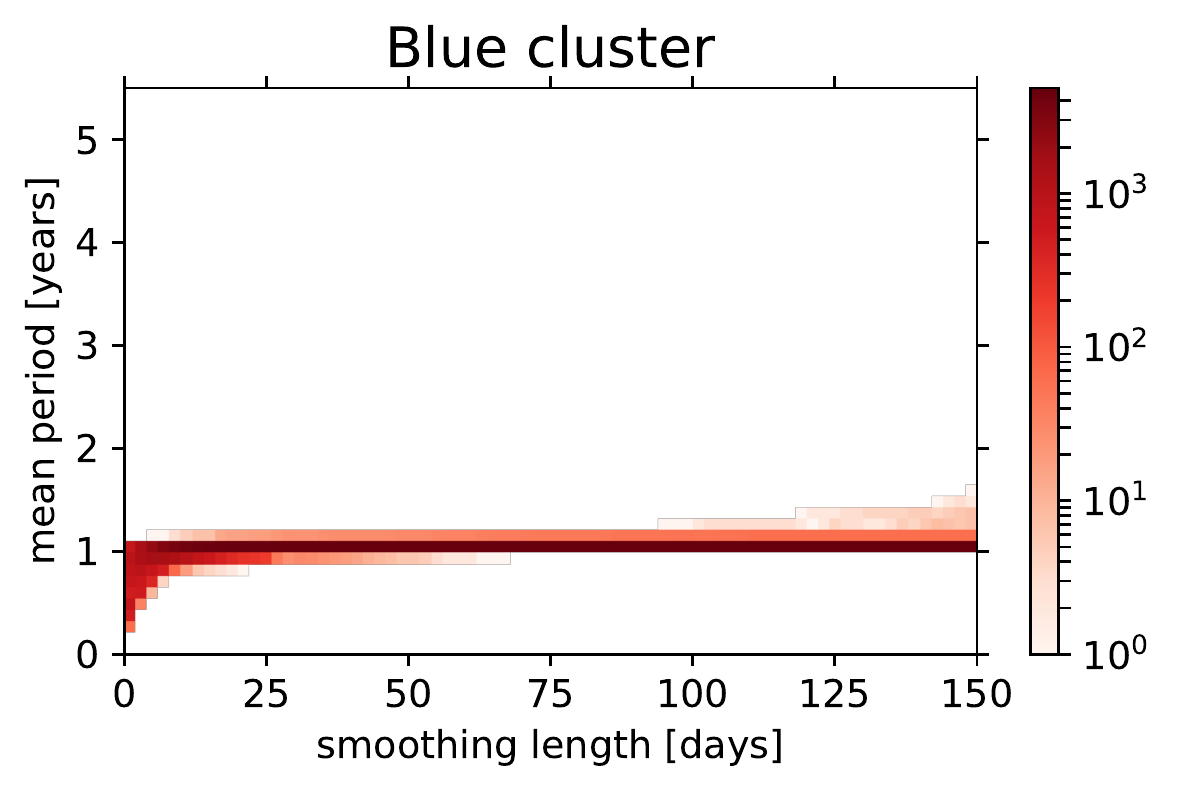}
		\includegraphics[width=0.49\columnwidth]{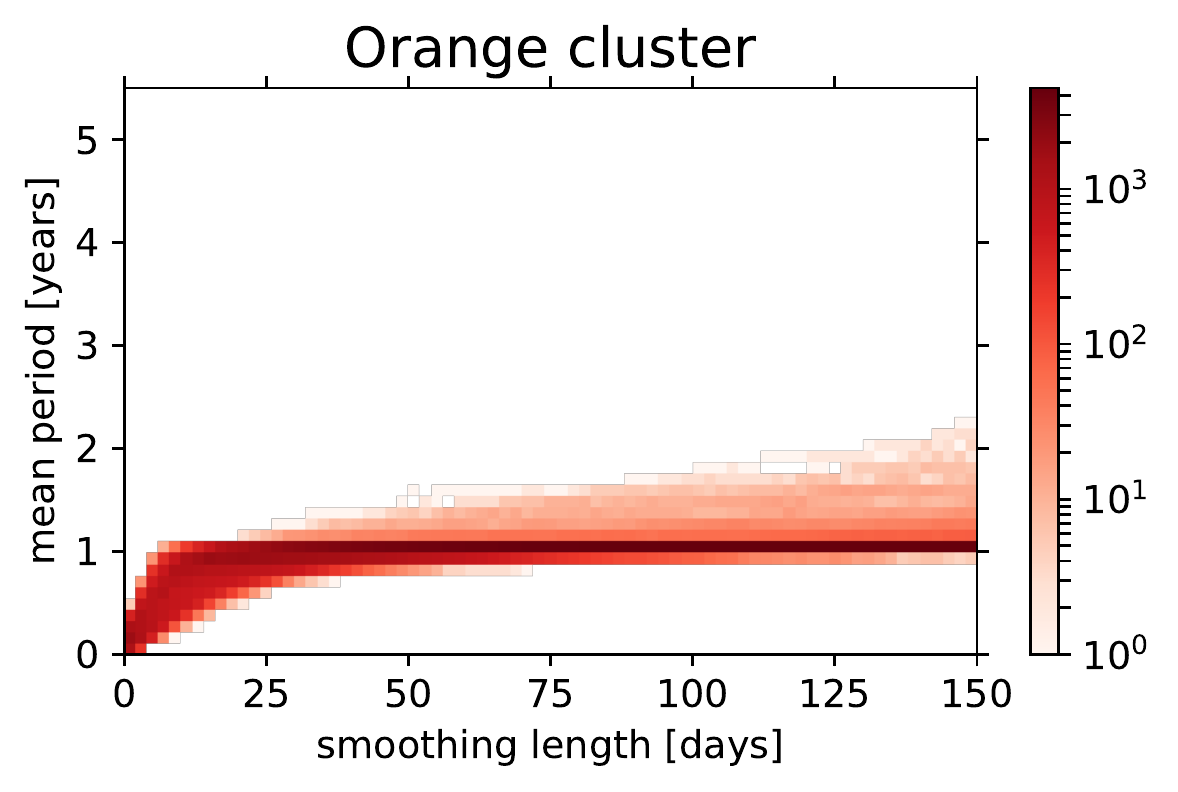}
		\\
		\includegraphics[width=0.49\columnwidth]{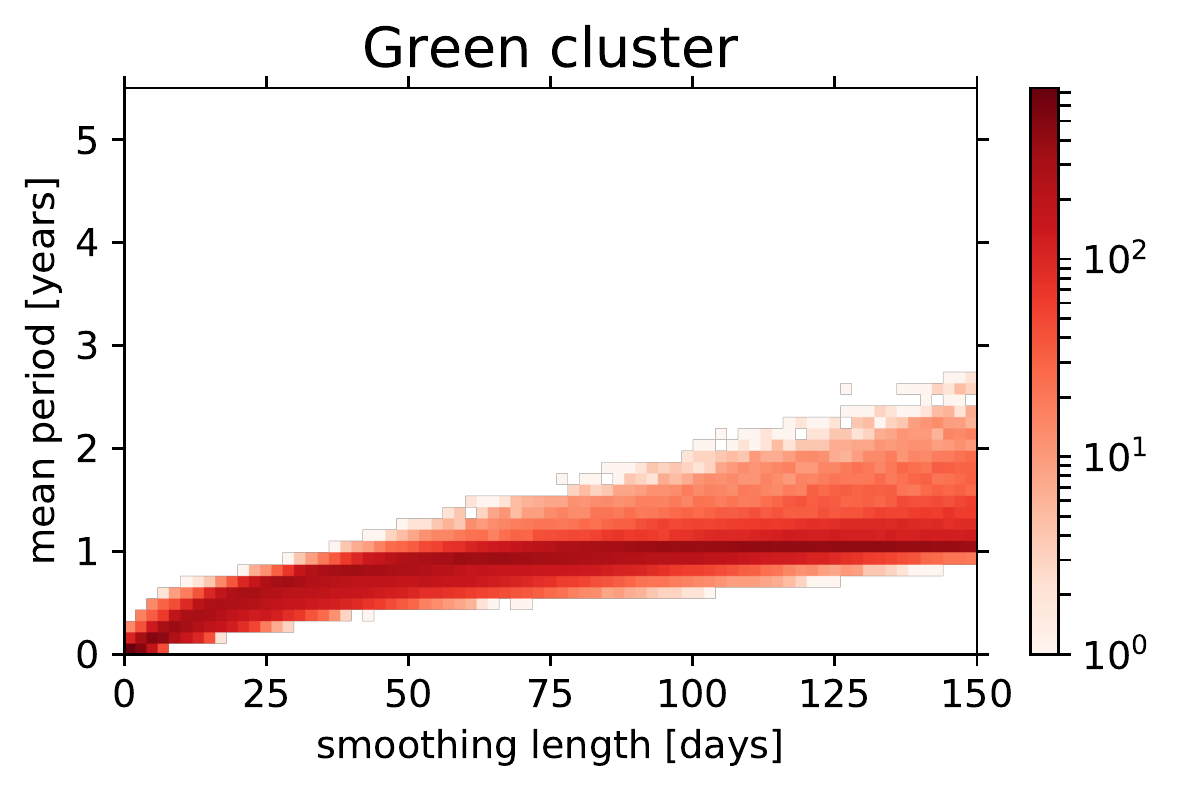}		
		\includegraphics[width=0.49\columnwidth]{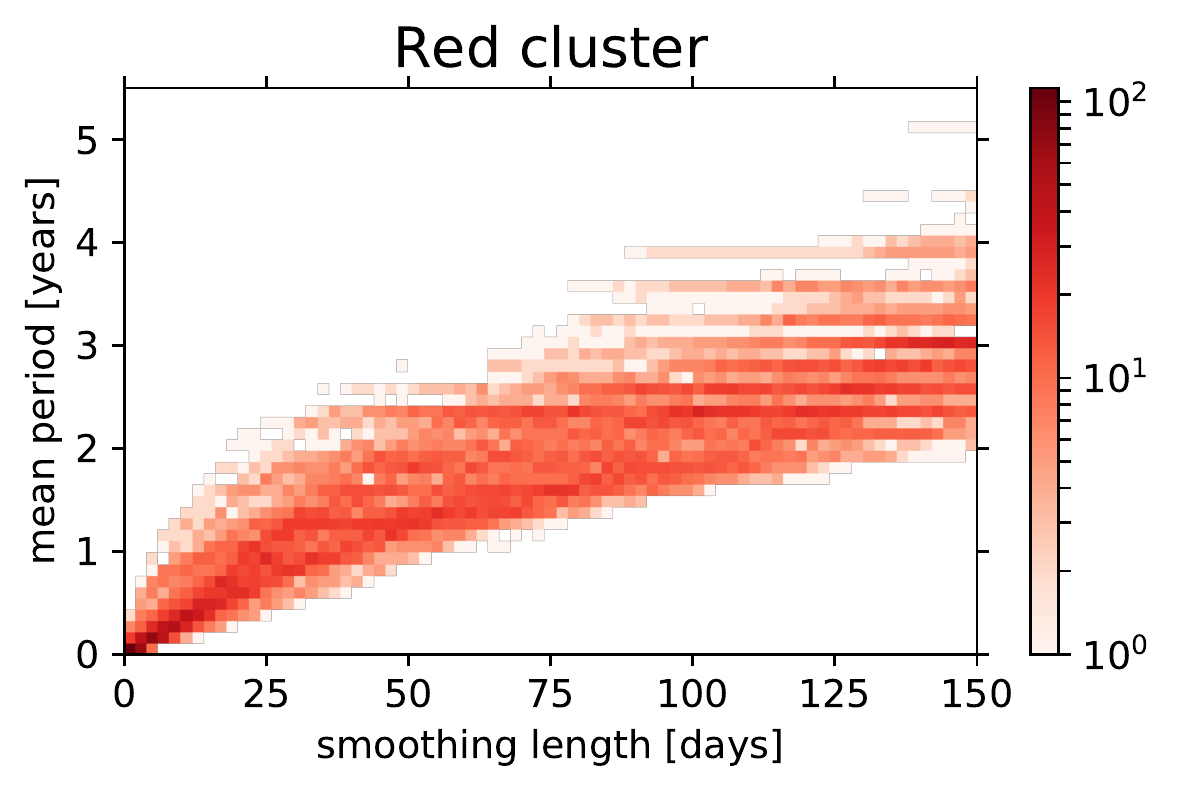}	
		\caption{Variation of $\overline{T}$ with $\tau$ in each cluster, represented as a 2D histogram.
		    The colour code indicates the number of sites in each bin; white represents empty bins.}
		\label{fig:map:4clusters_2}
	\end{figure}
	To further demonstrate that the analysis of the variation of $\overline{T}$ with $\tau$ indeed extracts meaningful information, we use the method of k-means clustering to classify the sites into a given number of clusters. Specifically, for each site we take as features the values of $\overline{T}$ for $\tau = 1, 9, 29, 49, \ldots, 149$ days. Then, we use the k-means algorithm to classify the 10512 geographical sites into $n$ clusters based on these features.
	In Figs. \ref{fig:map:4clusters} and \ref{fig:map:4clusters_2} we report the results for $n = 4$ (consistent results where obtained with larger $n$, the analysis is in progress and will be reported elsewhere). It is interesting to note that the map has similarities with the upper left panel of Fig.~\ref{fig:maps-avgperiod}.
	
	In particular, the blue cluster characterises regions dominated by the seasonal cycle and large temperature variations (see the regular site).
	These include the north extratropical land masses and storm track regions over the Pacific and Atlantic oceans.
	In the southern hemisphere, this cluster characterises the extratropical continents and subtropical oceans, but most of the extratropics are in the orange cluster.
	This latter cluster characterises regions with faster dynamics that are dominated by the annual cycle only after smoothing with $\tau > 20$ days, which may reflect the importance of subseasonal variability in the southern hemisphere.
	
	The tropical band is dominated by the green cluster, which is composed by regions of low temperature variability and whose spatial structure is closely related to the mean rainfall pattern. 
	The green cluster characterises regions where no plateau is found when increasing the smoothing length.
	The largest green region (where the irregular site is located) is the western Pacific warm pool, which presents temperature variability that is strongly tied to convection on short time scales and thus shows up in this study as with irregular behaviour.
	
	Finally, the red cluster characterises the central Pacific, the region that for $\tau > 30$ days has the slowest dynamics (as can be seen in Fig.~\ref{fig:maps-avgperiod}).
	Note that the core of the cold tongue region has a strong annual cycle and therefore it is within the blue cluster.
	The red cluster marks a transition region between the strong wind-driven equatorial cold tongue dynamics and the weak and thermodynamic-driven variability of the western Pacific warm pool.
	It is a region with relatively weak annual cycle and influenced by El Niño and the QBO and thus shows slow dynamics (see El Niño and QBO sites). 
	The quasi-regular and double-period sites are located in cluster borders because they are monsoon regions and thus rainfall and temperature variations are strongly related during the summertime, giving rise to large deviations from the annual cycle.
	Similar behaviour is expected in south-east Asia and central America and subtropical South America.

	\section{Conclusions}
	In summary, we have presented a novel method for extracting information from complex oscillatory signals.
	Using SAT time series, we have shown that Hilbert phase analysis combined with temporal averaging allows to extract different oscillatory modes, and by using a machine learning algorithm it provides a novel way to classify different types of SAT dynamics. 
	
	The proposed method is based on the analysis of the variation of the mean period of rotation of the Hilbert phase, $\overline{T}$, with the length, $\tau$, of the temporal average. We discovered that $\overline{T}$ vs.\ $\tau$ exhibits well-defined plateaus, which reveal hidden regularity of SAT dynamics. 
	
	We have shown that the plateau behaviour and the gradual variation of $\overline{T}$ with $\tau$ are not necessarily detected by Fourier analysis because, as $\tau$ increases, the frequency of the dominant peak in the power spectrum is either constant or changes abruptly.
	Thus, a main advantage of the proposed method is that it allows to detect the gradual emergence of temporal regularity that occurs as SAT variability is washed out.
	
	Using synthetic data generated with a simple model, we have tested and validated the proposed method: in the synthetic data the variation of $\overline{T}$ with $\tau$ is fully consistent with our knowledge of the model that generates the data.
	Moreover, the variation of $\overline{T}$ with $\tau$ in the synthetic data was found to be similar to that in real SAT data, which suggests that, in spite of the extremely complex atmospheric dynamics, the basic mechanisms needed for understanding our findings can be surprisingly simple.
	
	
	\begin{acknowledgements}
		We wish to acknowledge the support the LINC project (EU-FP7-289447) and the Spanish MINECO/FEDER (FIS2015-66503-C3-2-P).
		C.M.\ acknowledges the ICREA Academia (Generalitat de Catalunya);
		D.A.Z.\ acknowledges the FI scholarship of AGAUR (Generalitat de Catalunya);
		D.A.Z.\ also acknowledges the help received by Andrea La Rosa and Carlo Corsaro with the application of the k-means algorithm.
	\end{acknowledgements}
	
	\appendix
	
	\section{Overview of Hilbert transform}
	
	The Hilbert transform (HT) provides, for a real signal $x(t)$, an \emph{analytic signal} $h(t)$, from where an instantaneous amplitude and an instantaneous phase can be defined:
	\begin{equation}
	h(t) = x(t) + iy(t) = A(t) e^{i\varphi(t)}.
	\end{equation}
	Here $y(t)$ is the Hilbert transform of $x(t)$:
	\begin{equation}
	y(t) = H[x](t) = \frac{1}{\pi} \mathsf{P.V.}\int_{-\infty}^{+\infty}\frac{x(\tau)}{t-\tau}d\tau,
	\end{equation}
	where P.V. means principal value, and $A$ and $\varphi$ can be calculated as: $A(t) = \sqrt{\left[x(t)\right]^2 + \left[ y(t) \right]^2}$, and $\varphi(t) = \arctan [{y(t)}/{x(t)}]$.
	These series, obtained by HT, allow us to reconstruct the original series as $x(t) = A(t) \cos \varphi(t)$.
	
	As a first example, the Hilbert transform of the harmonic oscillation $x(t) = A \cos(\omega t)$ is $y(t) = A \sin(\omega t)$.
	In the complex plane,  $(x(t), y(t))$ represent the coordinates of a point that describes a circular trajectory of amplitude $A$ and phase $\omega t$.
	
\begin{figure*}[tbp]
		\centering
		\includegraphics[height=0.25\textwidth]{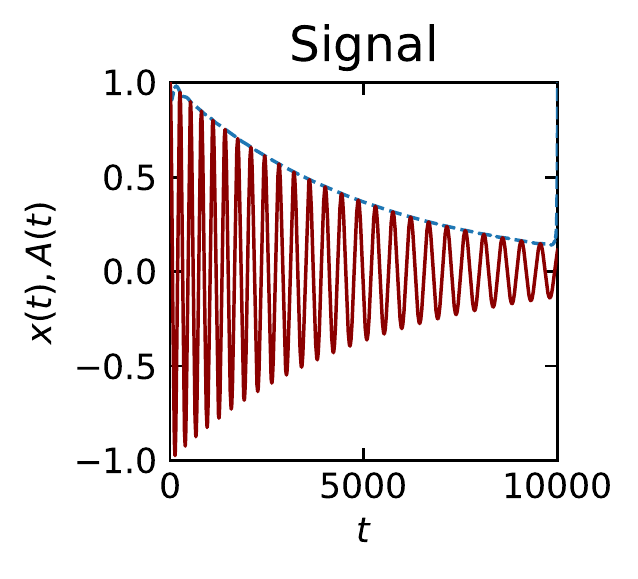}
		\qquad
		\includegraphics[height=0.25\textwidth]{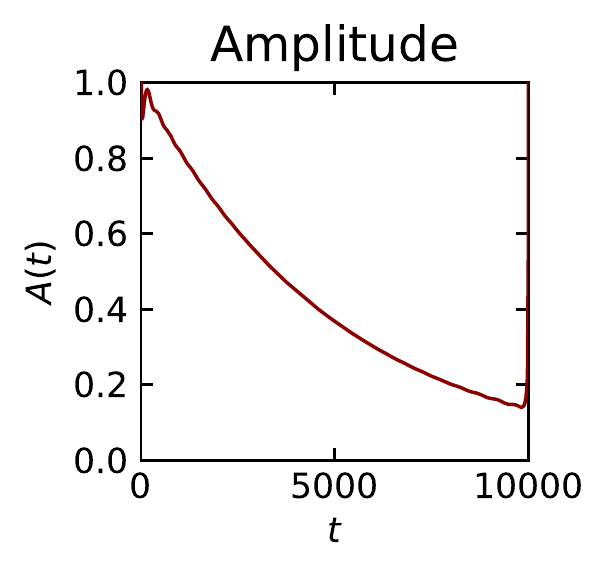}
		\qquad
		\includegraphics[height=0.25\textwidth]{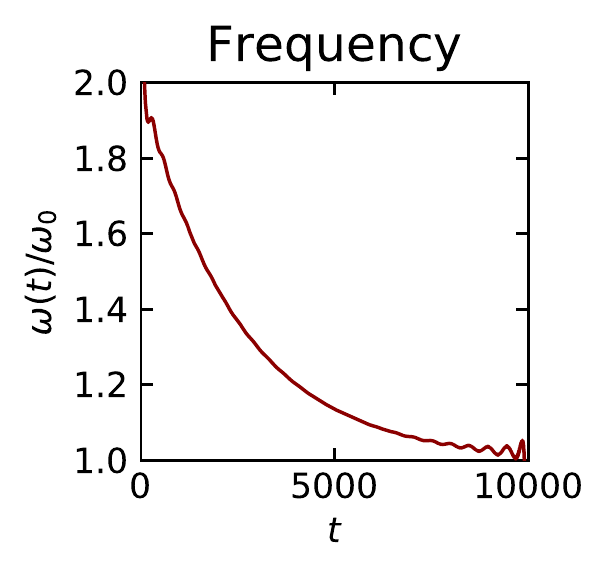}
		\caption{Hilbert analysis applied to the time series with time-varying amplitude and frequency, given by Eq.~\ref{eq:example}.
		The first panel shows the signal $x(t)$ (red line) and its Hilbert amplitude $A(t)$ (blue dashed line);
		the second panel shows only the Hilbert amplitude $A(t)$;
		the third panel shows the ratio between the instantaneous Hilbert frequency and $\omega_0$.}
		\label{fig:Hilbert-exponential}
	\end{figure*}

	As a second example, we calculate the instantaneous amplitude and frequency of the oscillation described by
	\begin{equation}
	x(t) = \mathrm{e}^{- \alpha t} \cos \left[ \left( 1 + \mathrm{e}^{-2 \alpha t} \right) \omega_0 t \right].
	\label{eq:example}
	\end{equation}
	Its length is $L = 10^5$.
	We choose $\omega_0 = {2 \pi}/{500}$ (which would describe an oscillation of length 500 in a harmonic oscillator) and $\alpha = 2/L$.
	The results are shown in Figure~\ref{fig:Hilbert-exponential}.
    We can see that the amplitude $A(t)$ is the exponentially decreasing envelope of the signal $x(t)$, given by the expression $\mathrm{e}^{- \alpha t}$, while the frequency $\omega(t)$ decreases according to $\left( 1 + \mathrm{e}^{-2 \alpha t} \right) \omega_0$.
    We also note that, near the extremes, $A(t)$ and $\omega(t)$ display an oscillatory behaviour that deviates from the analytical expressions that we have just given.

	\section{Comparison with other SAT reanalyses}
	\label{app:compar}
	To demonstrate the robustness of our findings, here we compare the results obtained from ERA-Interim with those obtained from NCEP Reanalysis 2~\cite{bib:NCEP-Rean2}, which has daily resolution and a spatial resolution of $192 \times 94$ ($1.875^\circ$ in longitude and approximately $1.9^\circ$ in latitude).
	We analyse the same period as in the main text (1981--2015).
	
	\begin{figure*}[htbp] 
		\centering
		\begin{tabular}[t]{rrr}
			{\includegraphics[height=0.25\textwidth]{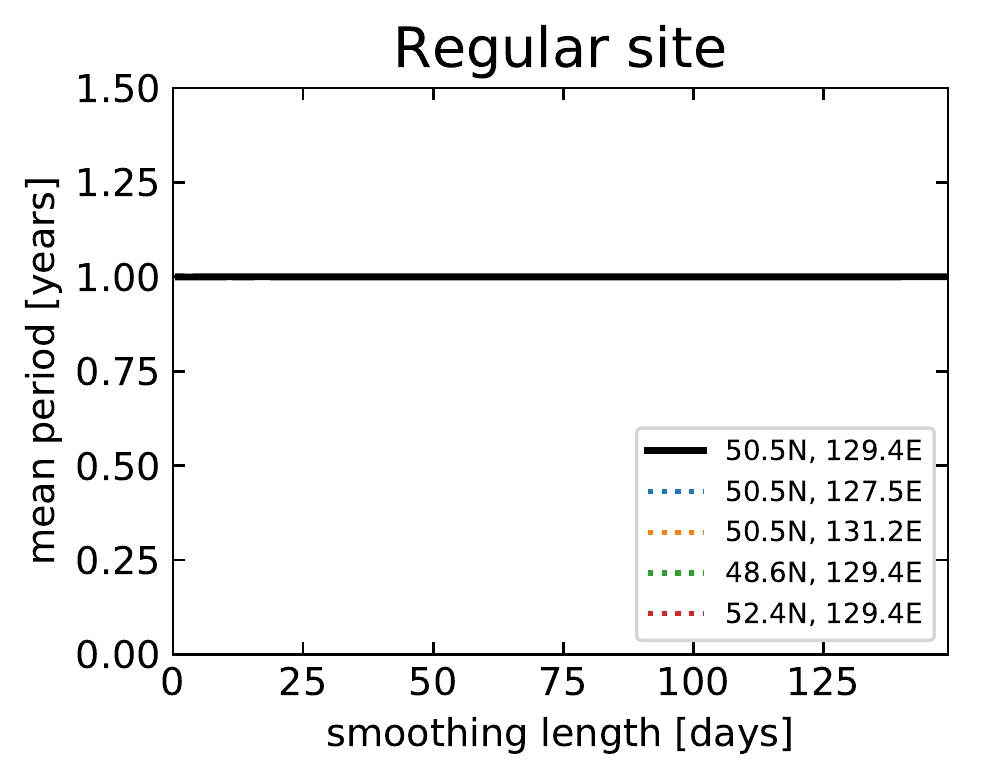}}
			&
			{\includegraphics[height=0.25\textwidth]{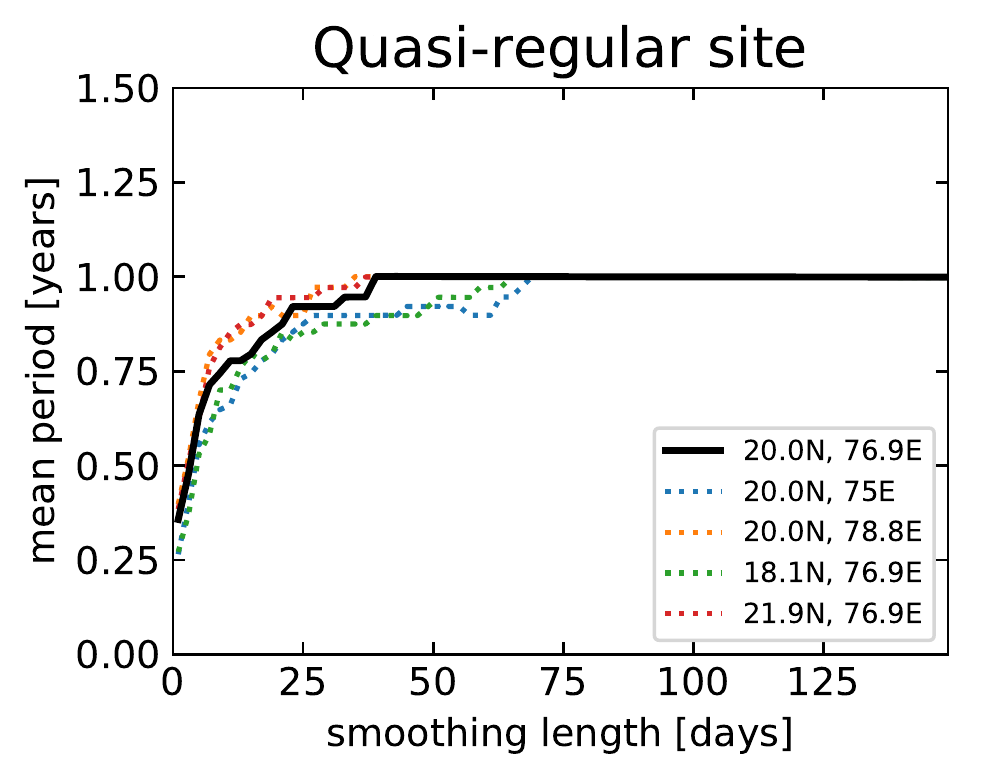}}
			&
			{\includegraphics[height=0.25\textwidth]{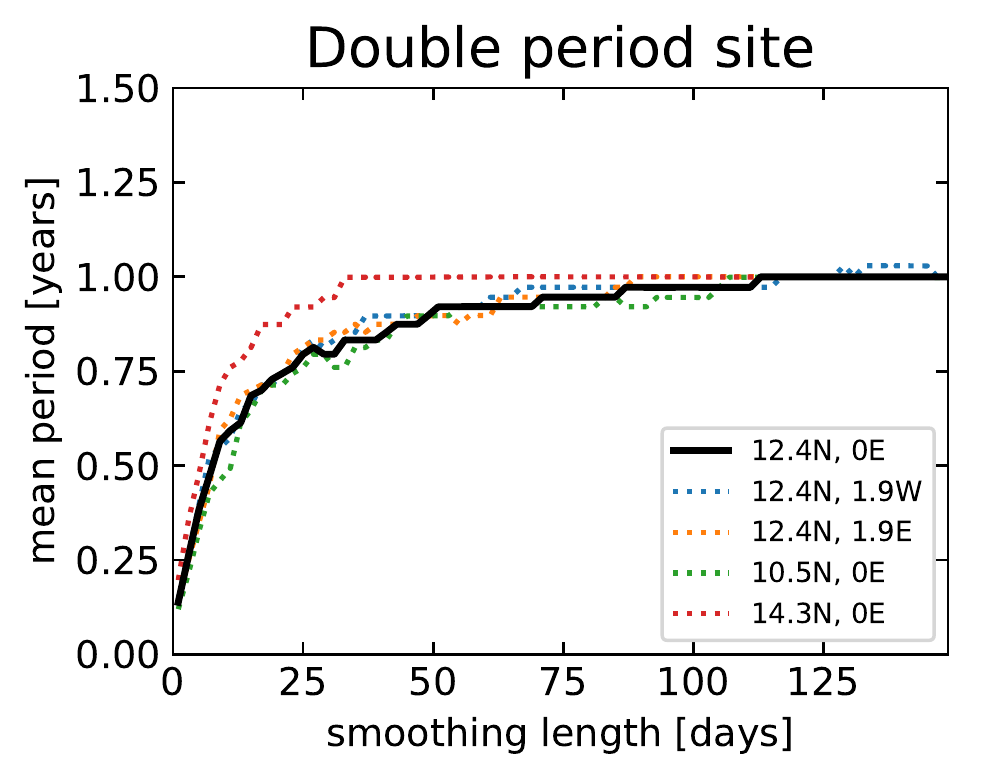}}
			\\
			{\includegraphics[height=0.25\textwidth]{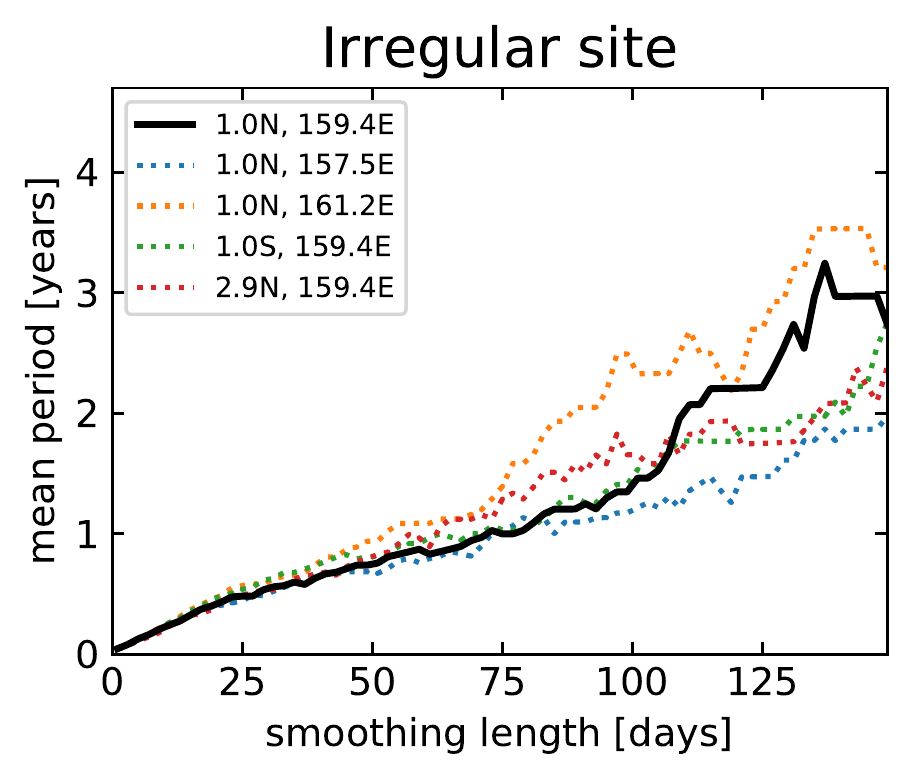}}
			&
			{\includegraphics[height=0.25\textwidth]{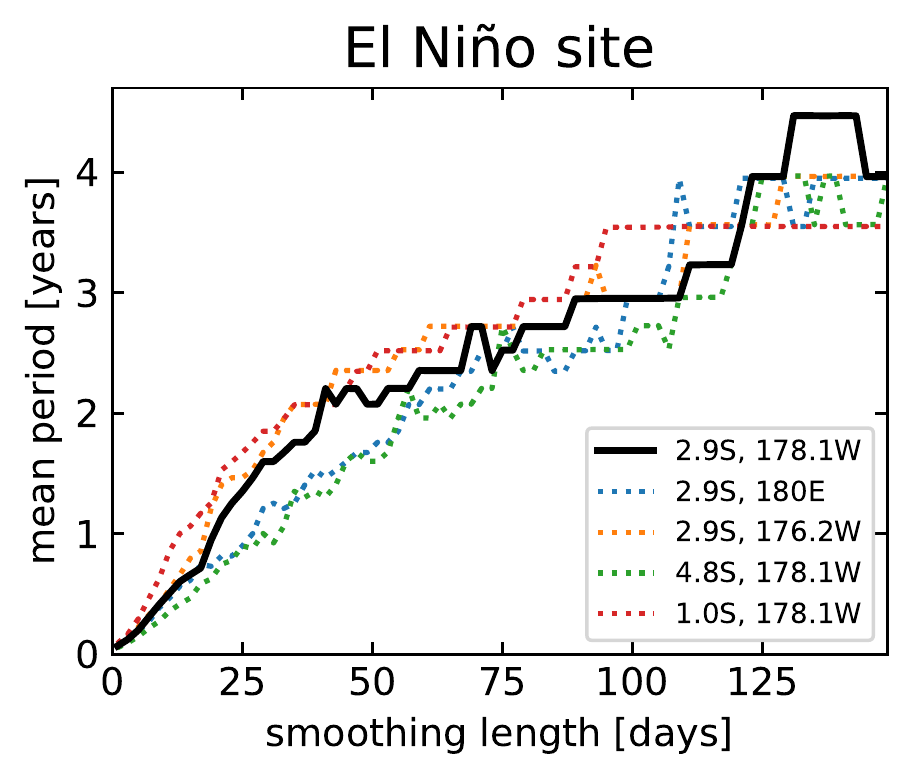}}
			&
			{\includegraphics[height=0.25\textwidth]{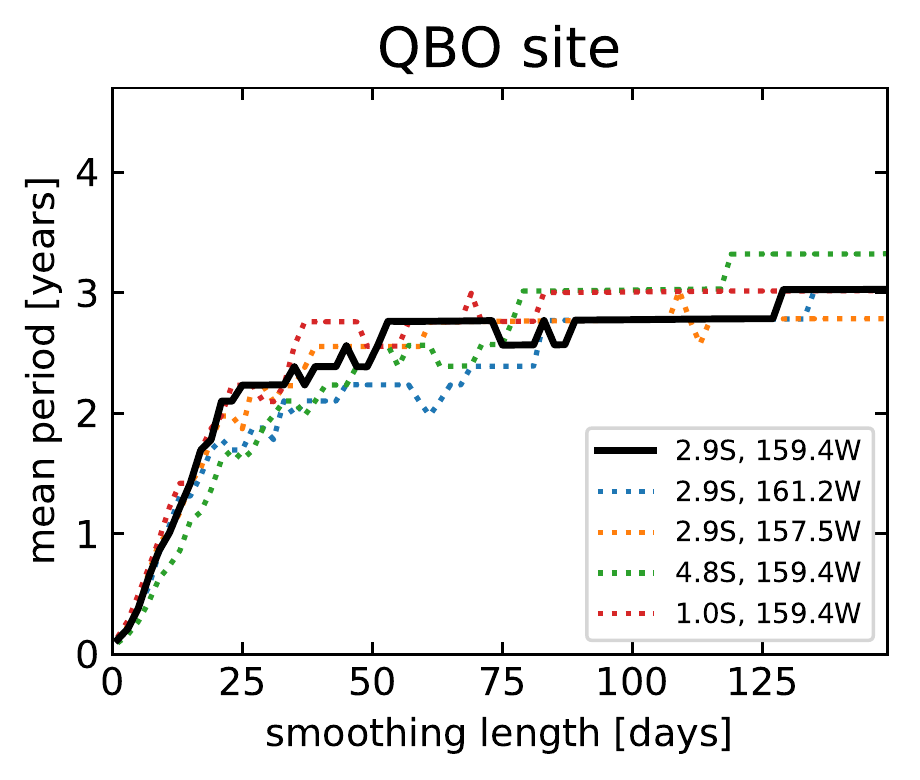}}
		\end{tabular}
		\caption{Mean period as a function of the smoothing length, for NCEP Reanalysis 2.}
		\label{fig:comparison-regular}
	\end{figure*}
	Figure~\ref{fig:comparison-regular} shows the variation of $\overline{T}$ with $\tau$, obtained from NCEP Reanalysis 2, for the six chosen geographical sites.
	Comparing with Fig.~\ref{fig:periodevol}, a good agreement is observed between the two reanalyses.
	There are some differences in the sites near to equator (specifically, double period site and El Niño site) that may be due to the fact that NCEP and ERA-Interim reanalysis are not in the same spatial grid, and the gradients in behaviour can be very large.
	In other words, in this region small changes of position can give very different results.
	
	\begin{figure}[htbp] 
		\centering
		\includegraphics[width=\columnwidth]{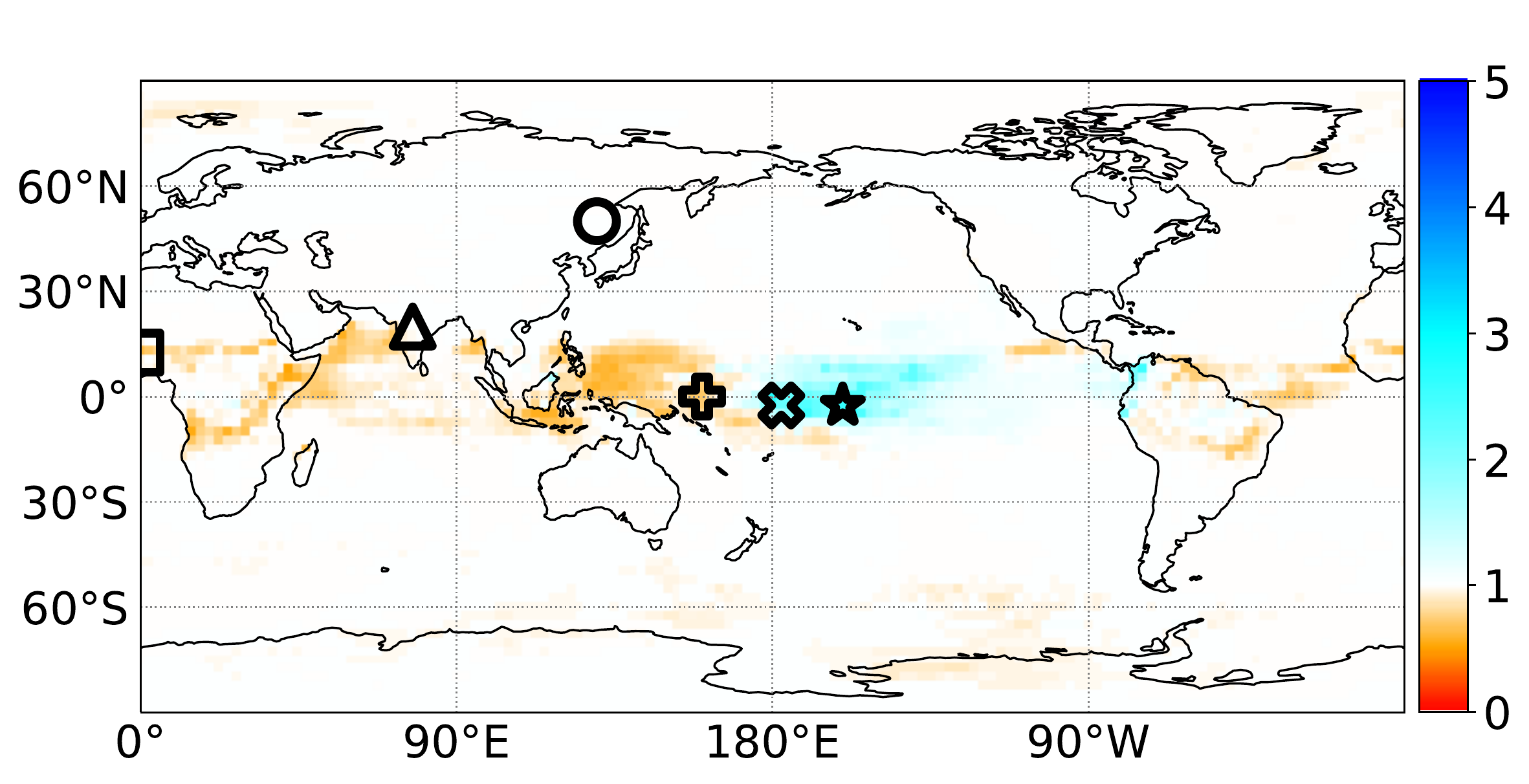}
		\caption{
			The colour map displays $\overline{T}$ (measured in years) computed from monthly SAT time series.}
		\label{fig:map-period-monthly}
	\end{figure}
	Finally, by considering ERA-Interim reanalysis with monthly resolution, we test the influence of the temporal resolution.
	In Figure~\ref{fig:map-period-monthly} we can see that the mean period obtained with monthly resolution and no smoothing is consistent with the results presented in Figure~\ref{fig:maps-avgperiod} and obtained with daily resolution.
	Specifically, we note that this map obtained from monthly SAT data looks qualitatively as an intermediate case between the maps obtained from daily SAT data with $\tau = 31$ days and $\tau = 99$ days.

	\section{Modelling the synthetic series}
	In Section~\ref{analysis:synth} we applied Hilbert analysis to synthetic series and showed that it returns information which is fully consistent with our knowledge of the equation that generates the series.
	Here we also show that this synthetic model provides a simple way to understand and reproduce the variation of $\overline{T}$ with $\tau$ found in real SAT data.
	
	\begin{figure}[tbp] 
		\centering
		\begin{tabular}[t]{rrr}
			\includegraphics[height=0.3\columnwidth]{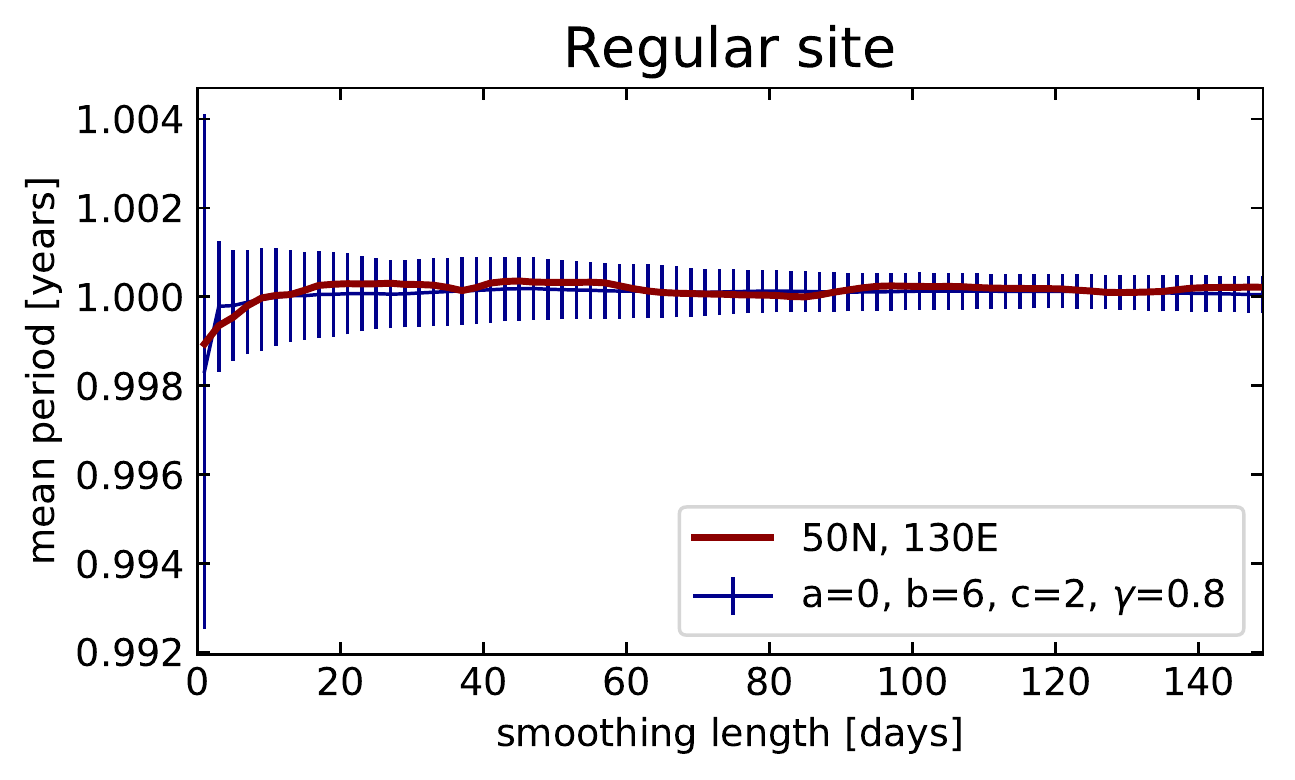}
			&
			\includegraphics[height=0.3\columnwidth]{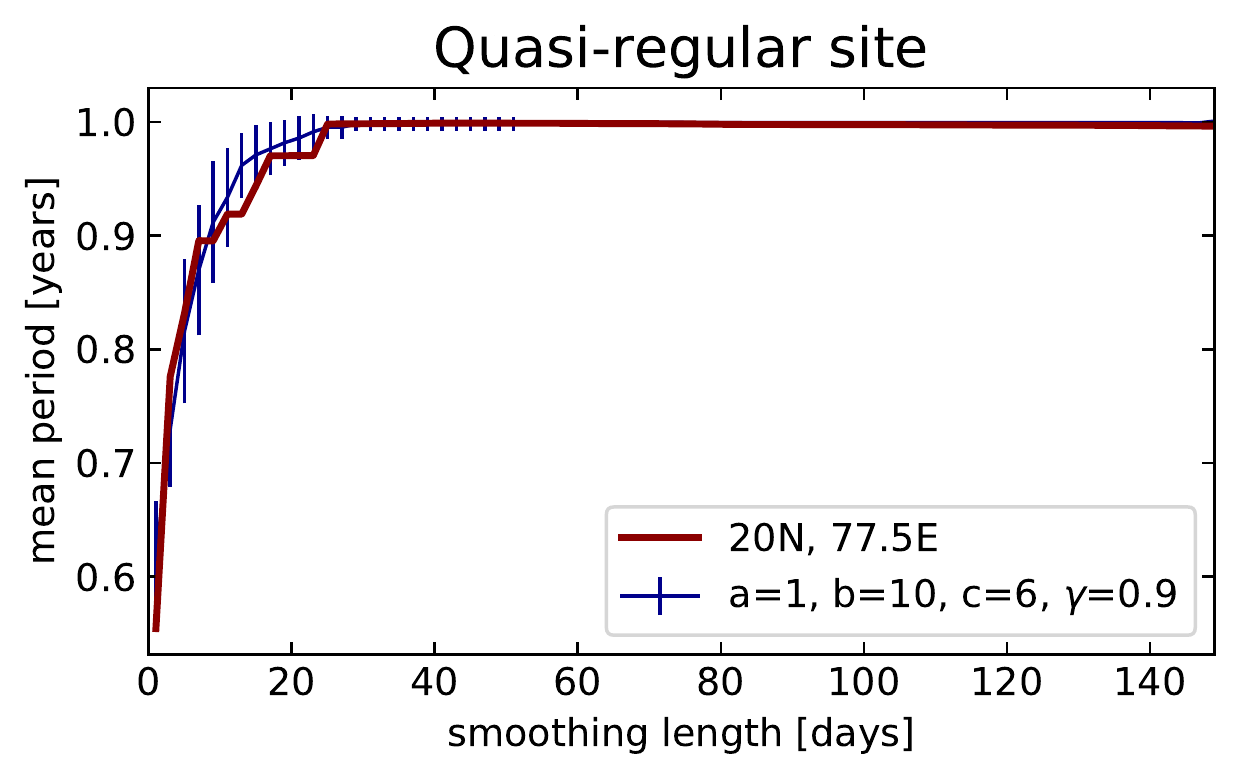}
			\\
			\includegraphics[height=0.3\columnwidth]{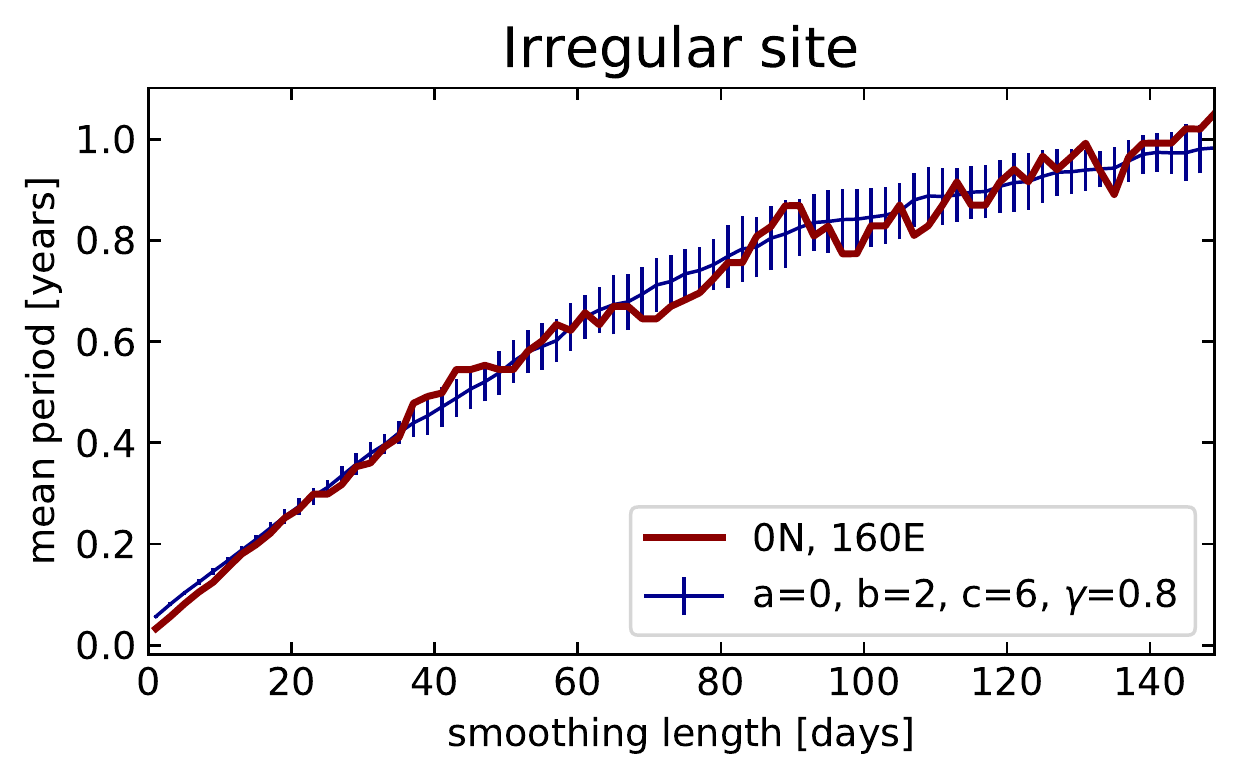}
			&
			\includegraphics[height=0.3\columnwidth]{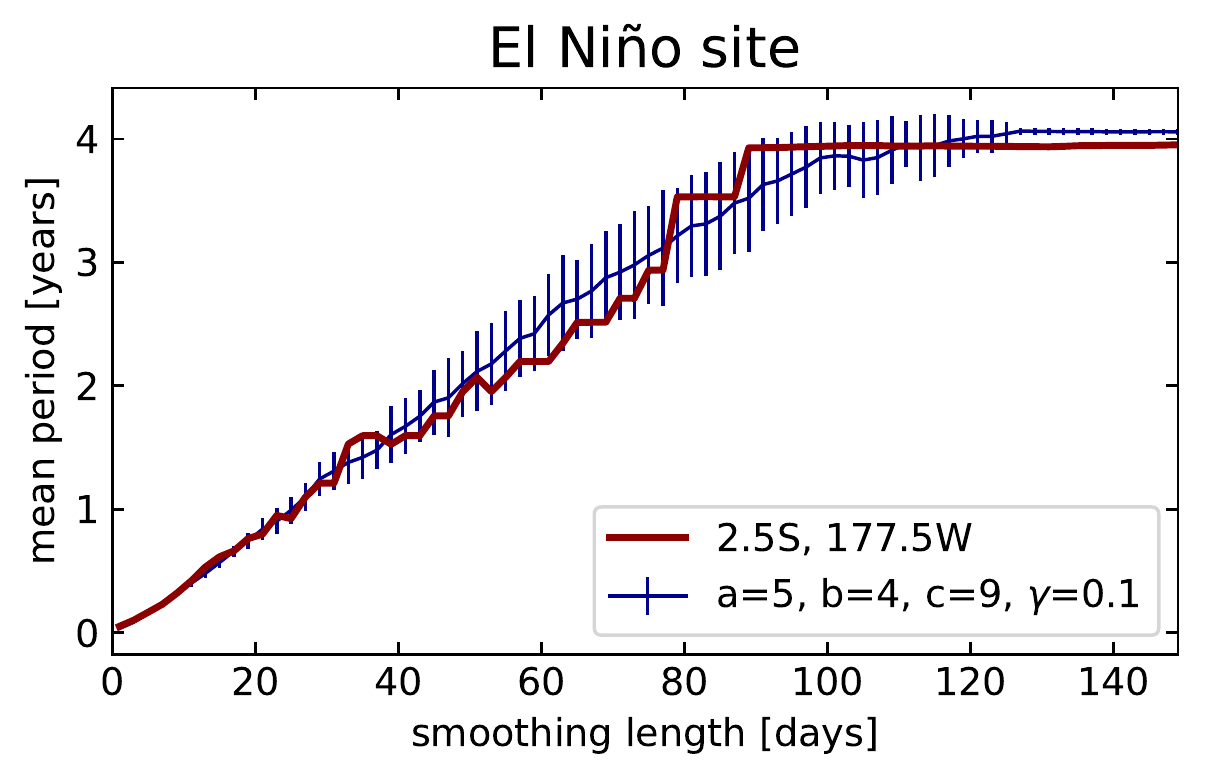}
		\end{tabular}
		\caption{Mean period $\overline{T}$ as a function of the smoothing length $\tau$.
			Red colour indicates results from the SAT series of an actual geographical site, while blue colour indicates results from the synthetic series that best fits to the results from the SAT series.
			We make 20 realisations of the synthetic series and calculate the average values of $\overline{T}$ as a function of $\tau$.
			The error bars represent the standard deviations.
			In the legend of each panel, we show the coordinates of the geographical site and the parameter values that best fit the period variation.}
		\label{fig:comp-synth}
	\end{figure}
	Figure~\ref{fig:comp-synth} displays the variation of $\overline{T}$ and compares the results of real and synthetic time series.
	For each of the four represented geographical sites, we search for the synthetic series that best fits the variation of $\overline{T}$ with $\tau$ obtained from the real SAT series.
	To find it, we vary $a,b,c$ from 0 to 10 with steps of 1, while $\gamma$ varies from 0 to 0.9 with steps of 0.1.
	For each choice of the parameter set, we make 20 realisations of the synthetic series and we calculate the average values of $\overline{T}$ as a function of $\tau$.
	Then, we calculate the sum of the squared distances between the variation of $\overline{T}$ given by the real series and the one given by the synthetic series.
	We select as the best fit the parameter set that minimises this sum.

	We can see that the regular site is fitted by our model without slow cycle and with low noise ($a=0, b=6, c=2$). The quasi-regular site is fitted when the noise is higher, but still lower than the one-year cycle ($a=1, b=10, c=6$). The irregular site is fitted when the noise is higher that the one-year cycle ($a=0, b=2, c=6$). On the other hand, the El Niño site is fitted when the slow cycle and the one-year cycle have comparable amplitudes, and the noise is higher than the cycles' amplitudes ($a=5, b=4, c=9$).

	\nocite{*}
	
\end{document}